\documentclass[a4paper,11pt]{article}
\pdfoutput=1 

\usepackage{jheppub} 

\usepackage[T1]{fontenc} 

\usepackage{graphicx}
\usepackage{grffile}  
\usepackage{xspace}      
\usepackage{amsmath}
\usepackage{slashed}
\usepackage{soul}
\usepackage{subcaption}
\usepackage{color}
\definecolor{nicered}{rgb}{0.5,0.,0.}
\definecolor{nicegreen}{rgb}{0.,0.5,0.}
\definecolor{niceblue}{rgb}{0.,0.,0.5}

\newcommand{\Lagrangian[1]}{\mathcal{L}_{#1}}

\title{\boldmath Heavy neutrinos at future linear e$^{+}$e$^{-}$ colliders}

\author[a]{Krzysztof~M\k{e}ka{\l}a,}
\author[b]{J\"urgen Reuter\;}
\author[a]{and Aleksander Filip~\.Zarnecki\;}
\affiliation[a]{Faculty of Physics, University of Warsaw,
                Pasteura 5, 02-093 Warszawa, Poland}
\affiliation[b]{Deutsches Elektronen-Synchrotron DESY, Notkestr. 85, 22607 Hamburg, Germany}

\emailAdd{k.mekala@uw.edu.pl}
\emailAdd{juergen.reuter@desy.de}
\emailAdd{zarnecki@fuw.edu.pl}

\newcommand{\whizard}{\textsc{Whizard}\xspace}

\newcommand{\delphes}{\textsc{Delphes}\xspace}

\newcommand{\epem}{\ensuremath{\text{e}^+\text{e}^-}\xspace}

\abstract{
Neutrinos are among the most mysterious particles in nature. Their mass hierarchy and oscillations, as well as
their antiparticle properties, are being intensively studied in experiments around
the world. Moreover, in many models of physics beyond the Standard Model, the baryon
asymmetry or the dark matter density in the Universe are explained by
introducing new species of neutrinos. Among others, heavy neutrinos of Dirac or Majorana nature were proposed to solve open questions in High Energy Physics. Such neutrinos with masses above the electroweak (EW) scale could be
produced at future linear \epem colliders, like the Compact LInear
Collider (CLIC) or the International Linear Collider (ILC).

We studied the possibility of observing decays of heavy Dirac and Majorana neutrinos
in the $qq\ell$ final state with ILC running at 500\,GeV and 1\,TeV, and CLIC at 3\,TeV. The
analysis is based on the \whizard event generation and fast simulation
of detector response with \delphes. Neutrinos with masses from
200 GeV to 3.2 TeV were considered. We estimated the limits on the
production cross sections, interpreted them in terms of the neutrino-lepton coupling parameter $V_{\ell N}^{2}$ (effectively the neutrino mixing angle) and
compared them with current limits coming from the LHC running at 13
TeV, as well as the expected limits from future hadron
colliders. The limits for the future lepton colliders, 
extending down to the coupling values of $10^{-7} - 10^{-6}$,
are stricter than any other limit estimates published so far.
}

\keywords{Sterile or Heavy Neutrinos, Specific BSM Phenomenology, \epem experiments}

\graphicspath{ {./plots/} }

\begin{document} 

\maketitle
\flushbottom


\section{Introduction}

In several models of New Physics, some open problems of the Standard Model (SM), such as the baryon asymmetry in the universe, the flavour puzzle, or the nature of the dark matter (DM), are solved by introducing new species of neutrinos of either Dirac or Majorana nature (see e.g. \cite{Canetti:2012vf, Caputo:2018zky, Gninenko:2013tk}). A sector of sterile neutrinos connected to the SM by mixing with the SM neutrinos could exhibit additional CP violation needed to explain the baryon asymmetry in the universe. The lightest sterile neutrino could be stable or so long-lived that it constitutes a considerable amount of the major part of DM. The neutrino sector also plays a prominent role in models with both lepton-flavour violation and lepton-flavour non-universality, which could explain the recent LHC flavour anomalies~\cite{Azatov:2018kzb}. There are also proposed connections of the anomaly in the magnetic moment of the muon $(g-2)_\mu$ to the neutrino sector~\cite{Hisano:2001qz,Dutta:2020scq}. 
Different mechanisms can be considered for production of such Dirac or Majorana neutrinos with masses exceeding several GeV at existing or future high-energy colliders, e.g. via exchange of SM gauge bosons, additional $Z'$ or $W'$ bosons, or in decays of heavy new particles like leptoquarks. For scenarios when the production mechanism is via the weak force, lepton colliders seem to be the most suitable devices for the heavy neutrino searches. There are two distinct scenarios. "Light heavy neutrinos" with masses below the $Z$ mass can occur in decays of the $Z$ and $W$ boson, and the large luminosity of future $Z$ and electroweak factories (for recent studies cf. \cite{Ding:2019tqq, Shen:2022ffi}) would give the best search limits, together with the high-luminosity phase of the LHC (HL-LHC). There is a small intermediate phase where the neutrino would be heavier than $W$ and $Z$, but lighter than the $H(125)$ Higgs boson. Then, it could occur in (invisible) Higgs decays, but it will be hard to distinguish them from e.g. Higgs portal models. As soon as the neutrino masses are
above the electroweak scale,  the heavy neutrinos can be produced at future linear $e^{+}e^{-}$ colliders, like the Compact Linear Collider (CLIC)~\cite{Linssen:2012hp} or the International Linear Collider (ILC)~\cite{Behnke:2013xla}. The signatures observable at lepton colliders have already been discussed in the literature (see e.g. \cite{delAguila:2005pin, delAguila:2005ssc, Saito:2010xj, Das:2012ze, Antusch:2016ejd, Banerjee:2015gca, Chakraborty:2018khw, Das:2018usr, Cai:2017mow}), but detailed, quantitative studies taking into account all relevant experimental effects have been missing so far.

Many different heavy neutrino production scenarios have been studied at the LHC.
For high masses of the new neutral lepton, above the EW boson masses, the highest sensitivity is expected for the heavy Majorana neutrino searches in the tri-lepton or same-sign di-lepton channels. 
Limits on the coupling parameter $V_{\ell N}^{2}$  extend down to about $10^{-5}$ for neutrino masses between 10 and 50\,GeV \cite{Sirunyan:2018mtv,ATLAS:2019kpx},
but are significantly weaker for masses above the Z boson mass scale. 
Limits on the new neutral lepton couplings for masses up to 50\,GeV can also be extracted from the analysis of W boson decays \cite{LHCb:2020wxx}.
Stronger limits, of the order of $10^{-6}$, were obtained from the search for long-lived particle decays (displaced vertex signature) \cite{ATLAS:2019kpx,CMS:2022fut}, which are however limited to low neutrino masses (below 10--15\,GeV).
Prospects for heavy Majorana neutrino searches were considered for future hadron colliders \cite{Pascoli:2018heg}, as well as electron-proton colliders \cite{Gu:2022muc}.

In this work, the possibility of observing the production and decays of heavy Dirac and Majorana neutrinos into the $qq\ell$ final state (corresponding to two measured jets and one lepton) at the ILC running at 500\,GeV and 1\,TeV, and the CLIC at 3\,TeV is studied. The analysis is based on \textsc{Whizard}~\cite{Moretti:2001zz,Kilian:2007gr} event generation and fast simulation of detector response with \textsc{Delphes}~\cite{deFavereau:2013fsa}. Dirac and Majorana neutrinos with masses from 200\,GeV to 3.2\,TeV are considered. We estimate limits on the production cross section and on the neutrino-lepton coupling using machine learning methods and compare them with current limits coming from the LHC running at 13\,TeV, as well as the expected future limits from hadron colliders. Beam-related effects and systematic uncertainties are included in the procedure. The expected limits obtained in our study are stronger than any other estimates published so far and exceed those for $pp$ machines by several orders of magnitude.

The paper is structured as follows: in Section 2, our model setup and simulation framework are described; in Section 3, we present our analysis procedure. Results are discussed in Section 4 and the most important features of the work and prospects of the analysis are summarised in Section 5.

\section{Model setup and simulation framework}

\subsection{Model setup}

There is a vast theory space of models of sterile neutrinos and extended neutrino sectors, regarding which pending problem of the SM is specifically addressed by them: they allow to introduce new sources of CP violation needed for leptogenesis or baryogenesis~\cite{Pilaftsis:1998pd, Buchmuller:2003gz, Drewes:2016gmt, Abada:2018oly, Barrow:2022gsu}, they introduce candidates for (cold) dark matter~\cite{Asaka:2005pn, Shaposhnikov:2008pf, Boyarsky:2018tvu} and they might play a role in the flavor puzzle~\cite{Baek:2020ovw}. Depending on whether they are embedded in extended gauge sectors, like e.g. in left-right symmetric models or Grand Unified Theories (GUTs), there will be additional gauge bosons above the electroweak scale in the multi TeV or not. For this study on the sensitivity reach of future high-energy lepton colliders, we stay mostly model-independent and assume that -- although there are up to three different heavy neutrino flavors -- only one of them is kinematically accessible at the studied colliders. No additional gauge bosons at any energy scale are assumed. The only interaction of the new neutrinos with the SM is through mixing effects, which come from a non-diagonal mass matrix between the electroweak doublet neutrinos and sterile neutrinos. Hence,
in this work, we focus on the Phenomenological Type I
Seesaw mechanism~\cite{delAguila:2008cj, Atre:2009rg}, implemented within the \textit{HeavyN} model~\cite{HeavyN} with Majorana~\cite{Alva:2014gxa, Degrande:2016aje} and Dirac~\cite{Pascoli:2018heg} neutrinos, an effective extension of the Standard Model introducing three flavours of right-handed neutrinos (denoted as $N_1$, $N_2$ and $N_3$) which are singlets under the SM gauge groups. 
\begin{figure}
    \centering
    \begin{subfigure}[b]{0.27\textwidth}
         \centering
         \includegraphics[width=\textwidth]{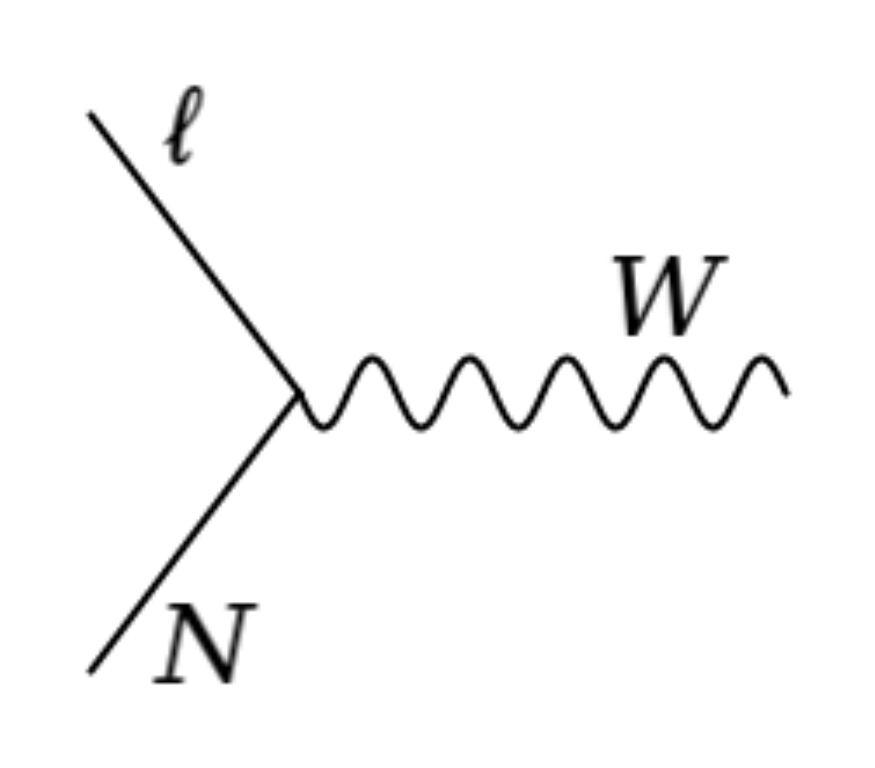}
     \end{subfigure}
     \begin{subfigure}[b]{0.27\textwidth}
         \centering
         \includegraphics[width=\textwidth]{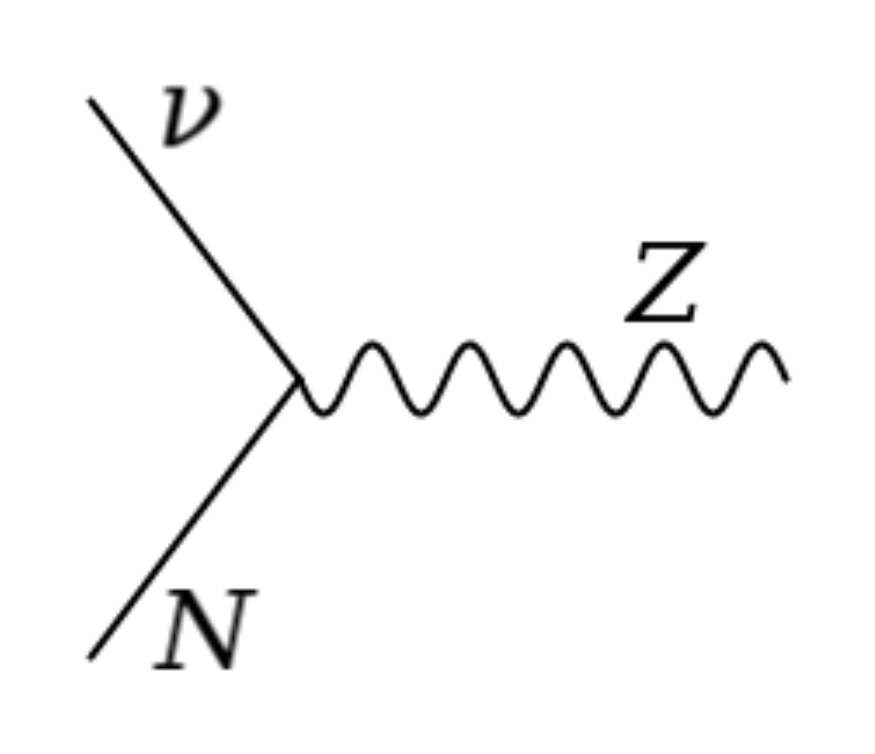}
     \end{subfigure}
     \begin{subfigure}[b]{0.27\textwidth}
         \centering
         \includegraphics[width=\textwidth]{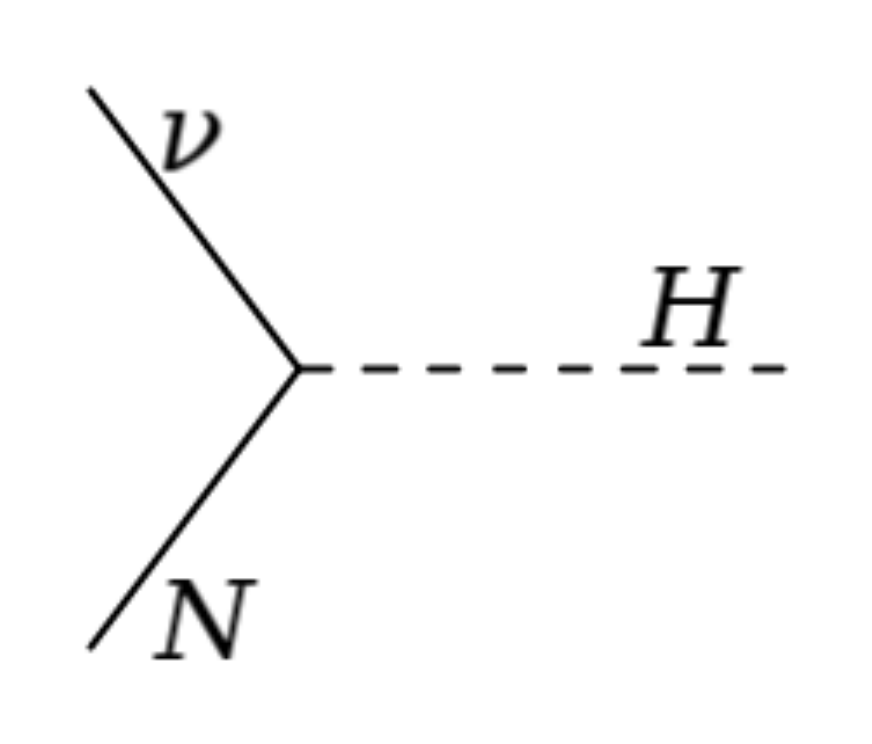}
     \end{subfigure}
    \caption{Extra vertices in the \textit{HeavyN} model: interactions of heavy neutrinos with $W$, $Z$ or Higgs bosons}
    \label{fig:vertices}
\end{figure}
The Lagrangian of the model is given by:
\begin{equation}
\Lagrangian[] 
= \Lagrangian[SM] + \Lagrangian[N] + \Lagrangian[WN\ell] + \Lagrangian[ZN\nu] + \Lagrangian[HN\nu]
\end{equation}
where $\Lagrangian[N]$ is a sum of kinetic and mass terms for heavy neutrinos (note that we use 4-spinor notation in all cases, which combines terms with spinors of dotted and undotted indices):
\begin{equation}
\Lagrangian[N] = \xi_\nu \cdot  \left(  \bar{N}_{k}i\slashed{\partial}N_{k} - m_{N_{k}}\bar{N}_{k}N_{k} \right) \qquad\textnormal{for}\; k = 1,\,2,\,3,
\end{equation}
with an overall factor $\xi_\nu = 1$ for the Dirac neutrino and $\xi_\nu = \frac{1}{2}$ for the Majorana neutrino scenarios.
$\Lagrangian[WN\ell]$ corresponds to neutrino interactions with a $W$ boson:
\begin{equation}
\Lagrangian[WN\ell] = - \frac{g}{\sqrt{2}}W^{+}_{\mu} \sum^{3}_{k=1}\sum^{\tau}_{l=e}\bar{N}_{k}V^{*}_{lk}\gamma^{\mu}P_{L}\ell^{-} + \textnormal{ h.c.},
\end{equation}
$\Lagrangian[ZN\nu]$ to interactions with a $Z$ boson:
\begin{equation}
\Lagrangian[ZN\nu] = - \frac{g}{2\cos\theta_{W}}Z_{\mu} \sum^{3}_{k=1}\sum^{\tau}_{l=e}\bar{N}_{k}V^{*}_{lk}\gamma^{\mu}P_{L}\nu_{l} + \textnormal{ h.c.},
\end{equation}
and $\Lagrangian[HN\nu]$ to interactions with a Higgs boson:
\begin{equation}
\Lagrangian[HN\nu] = - \frac{gm_{N}}{2M_{W}}h \sum^{3}_{k=1}\sum^{\tau}_{l=e}\bar{N}_{k}V^{*}_{lk}P_{L}\nu_{l} + \textnormal{ h.c.}
\end{equation}
The vertices involving $N$ introduced by the model are shown in Figure \ref{fig:vertices}.

The model is described in \textsc{FeynRules}~\cite{Christensen:2008py,Alloul:2013bka}, the \textsc{Mathematica} package to calculate Feynman rules associated with the Lagrangian of a given model. The output is stored in the \textsc{UFO} format~\cite{Degrande:2011ua}, the model format for automatized matrix element generators. The UFO library used in the analysis contains 12 free parameters in addition to the SM parameters:
\begin{itemize}
\setlength\itemsep{-0.2em}
    \item three masses of the heavy neutrinos: m$_{N_1}$, m$_{N_2}$ and m$_{N_3}$,
    \item nine real\footnote{As the parameters are chosen real, no $CP$ violation beyond the SM appears.} mixing parameters $V_{lk}$, where $l = e, \mu, \tau$ and $k = N1, N2, N3$.
\end{itemize}
There are also three widths of the heavy neutrinos ($\Gamma_{N1}$, $\Gamma_{N2}$ and $\Gamma_{N3}$) to be set.

\begin{figure}[tb]
    \centering
    \begin{subfigure}[b]{0.3\textwidth}
         \centering
         \includegraphics[width=\textwidth]{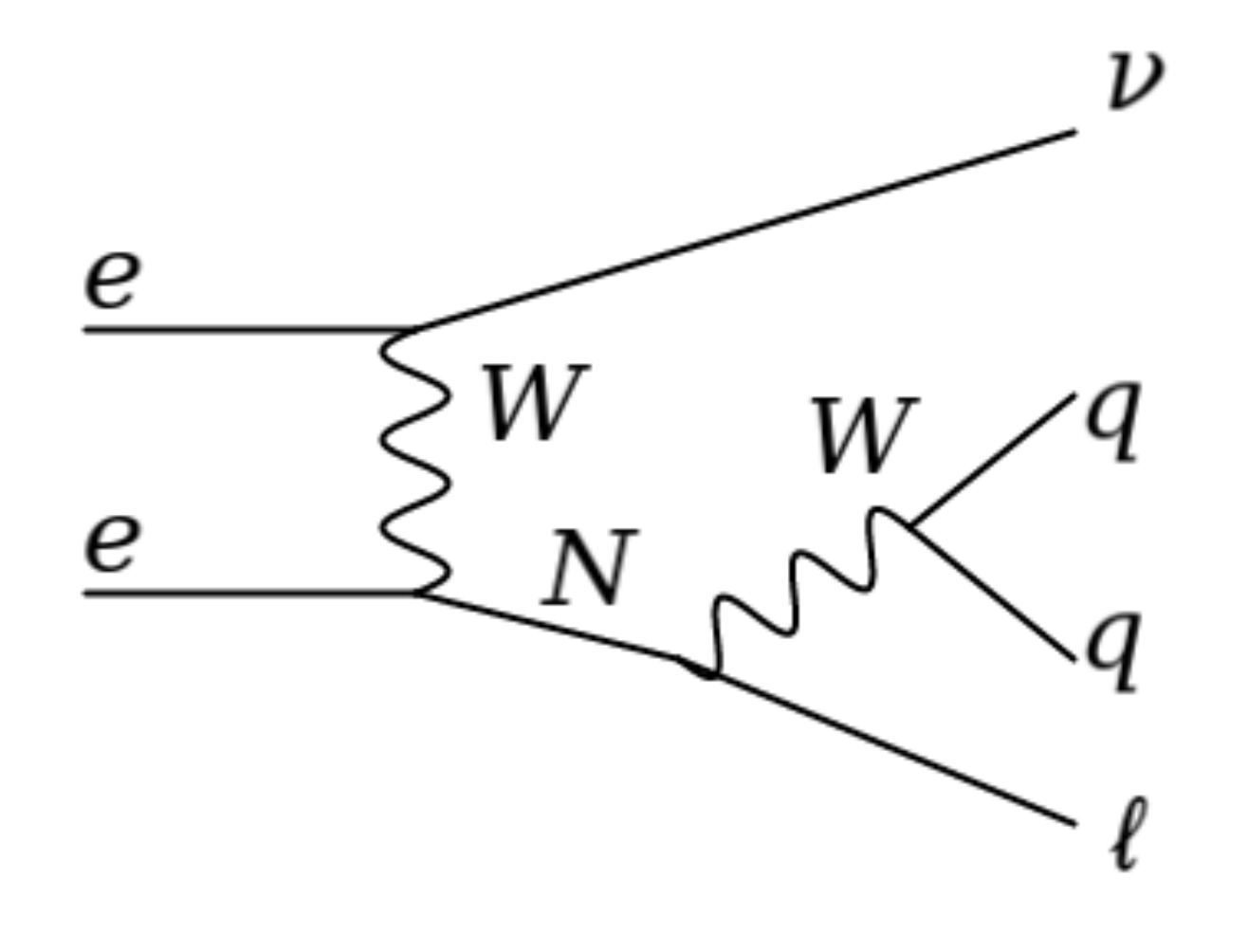}
     \end{subfigure}
     \begin{subfigure}[b]{0.3\textwidth}
         \centering
         \includegraphics[width=\textwidth]{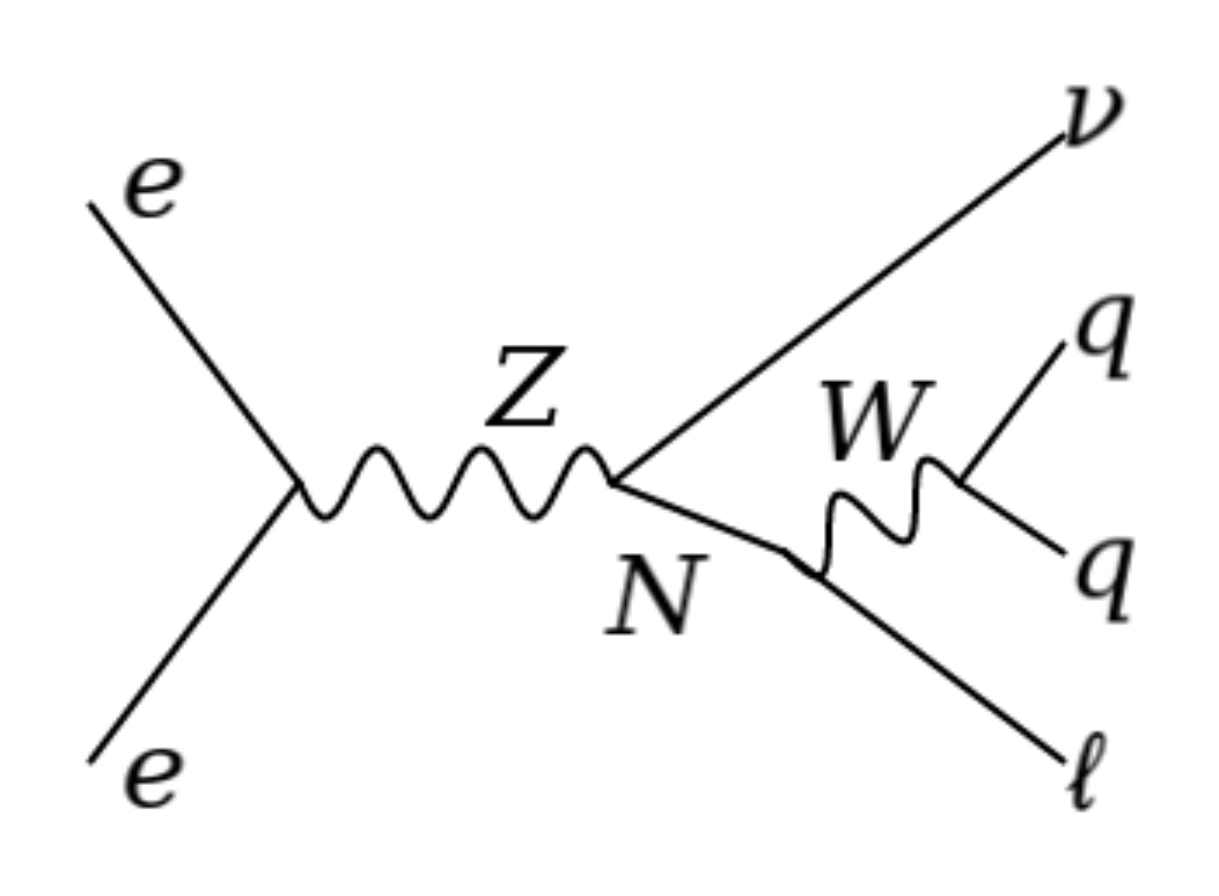}
     \end{subfigure}
    \caption{Feynman diagrams for heavy neutrino production at $e^+e^-$ colliders with the $qq\ell\nu$ signature}
    \label{fig:signal_diagram}
\end{figure}

For such neutrinos, there are many different signatures expected at future colliders~\cite{Antusch:2016ejd}. For $e^+e^-$ collisions, the dominant production channels are s-channel $Z$ production and t-channel $W$ exchange, resulting in the production of a light-heavy neutrino pair:
\[ e^+ e^- \rightarrow N \, \nu  \quad .\]
The $Z$ exchange process is dominant at the $Z$-pole (around the mass of the $Z$ boson), while for centre-of-mass energies above the $Z$-pole, the $W$ exchange contribution is more important. Analytic calculations show that the cross section for the production of a heavy-heavy neutrino pair is much lower and, hence, these processes are not considered in the analysis~\footnote{In addition, the vertex involving two heavy neutrinos was not implemented in the \textsc{FeynRules} model, and would have had to be added by hand.}.
In the parameter space considered, the heavy neutrino has a microscopic lifetime ($c\tau  \ll 1$ nm) so that no displaced vertices can be reconstructed and the products of its decays point back to the primary interaction point. Different final states are possible; however, in this paper, we focus on the $qq\ell\nu$ final state, corresponding, at the experimental level, to the $jj\ell$ signature. 
Example Feynman diagrams for the process are presented in Figure \ref{fig:signal_diagram}. The production process is dominated by the $W$ exchange for which only left-handed electrons and right-handed positrons contribute and thus, we decided to consider the corresponding beam polarisation settings.
Since the signal and the leading SM background channels depend on the polarisation in a similar way, such a choice allows for increasing the expected signal event number, keeping the signal-to-background ratio on the same order.
The following collider setups are considered: 
\begin{itemize}
\setlength\itemsep{-0.2em}
    \item ILC500 -- ILC running at 500 GeV, with an integrated luminosity of 1.6 ab$^{-1}$ and beam polarisation of $-$80\% for electrons and +30\% for positrons;
    \item ILC1000 -- ILC running at 1 TeV, with an integrated luminosity of 3.2 ab$^{-1}$ and beam polarisation of $-$80\% for electrons and +20\% for positrons;
    \item CLIC3000 -- CLIC running at 3 TeV, with an integrated luminosity of 4 ab$^{-1}$ and beam polarisation of $-$80\% for electrons (no polarisation for positrons).
\end{itemize}
For the dominant production channel, the above runs correspond to about 80\% of all data for ILC and 97\% for CLIC and the difference is mostly caused by the luminosity fraction assumed to be collected for each polarisation setup at those colliders.

\subsection{Event generation and benchmark scenarios}
The first step was to generate collision events using \textsc{Whizard}~\cite{Moretti:2001zz,Kilian:2007gr}. For the generation of SM backgrounds and Dirac neutrino samples, version 2.8.5 was used, while the simulation of the Majorana neutrino production was the first physics project using the new major version \textsc{Whizard 3.0.0}. The beam energy profile was parametrised with the \textsc{Circe2} subpackage within \textsc{Whizard}, parton showering and hadronisation were done with \textsc{Pythia~6}~\cite{Sjostrand:2006za}.

\begin{figure}[tb]
    \centering
    \includegraphics[width=0.8\textwidth]{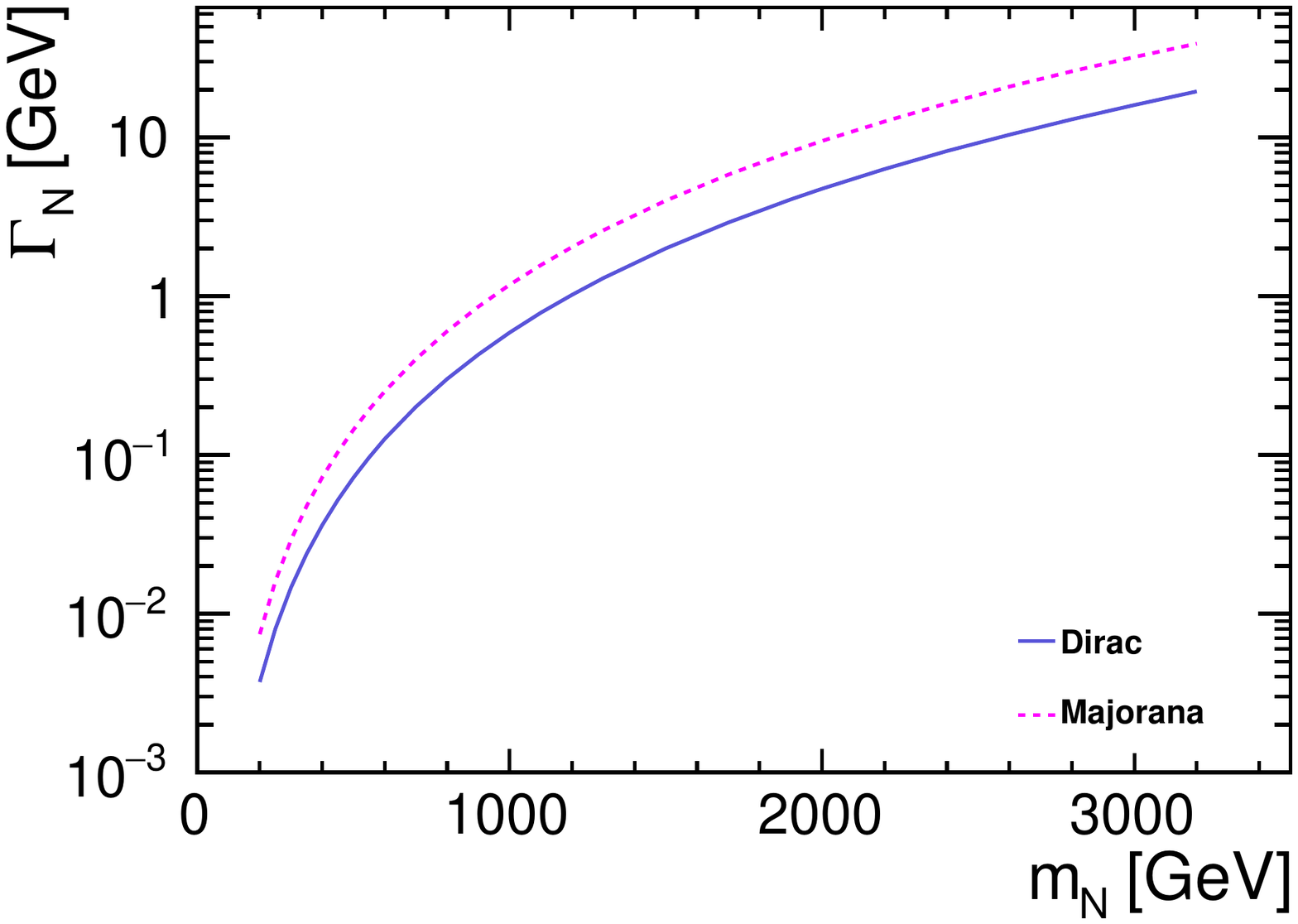}
    \caption{Calculated widths of the heavy Dirac (blue solid line) and Majorana (pink dashed line) neutrinos for the reference scenario ($V_{\ell N}^{2} = 0.0003$). For the Majorana case, the width is twice that for the Dirac case (cf. text).}
    \label{fig:widths}
\end{figure}

To generate signal events, the \textit{Dirac\_NLO} and \textit{Gen3Mass\_NLO} implementations of the \textit{HeavyN} model, described within the \textsc{FeynRules} model database, were used. To simplify the analysis, we assumed that only a single heavy neutrino is coupled to the Standard Model particles\footnote{The analysis is valid also in the case when a few heavy neutrinos exist but they have different masses and they are narrow enough such that there is no interference amongst them. On the other hand, if these states are nearly degenerate and are indistinguishable (e.g. regarding the flavors of their decay products), the experimental sensitivity to discover such particles increases~\cite{Chao:2009ef}.}. Therefore, for the simulation, the masses of $N_2$ and $N_3$ were set to 10\,TeV and their couplings to zero in the model. 
For the neutrino $N_1$ that is assumed to have non-vanishing coupling to the SM, and to which we refer from now on just as "heavy neutrino" or $N$, masses in the range 200-3200 GeV in steps of either 50, 100 or 200 GeV were considered as signal benchmark scenarios. For these scenarios, all the mixing parameters were set equal to $\sqrt{0.0003}$:
\begin{equation*}
|V_{eN}|^{2} = |V_{\mu N}|^{2} = |V_{\tau N}|^{2} = 0.0003 \equiv V_{\ell N}^{2}.
\end{equation*}
Widths of the heavy neutrino were calculated using \textsc{Whizard} and are in agreement with the values given in~\cite{Alva:2014gxa}. 
Because of the additional CP-conjugate final states, the widths for the Majorana case are twice as large as for the Dirac case.
The width values for the reference scenario are shown in Fig.~\ref{fig:widths} as a function of the heavy neutrino mass. One can observe that for the assumed coupling, the neutrino can be treated as a very narrow resonance, but the neutrino widths are not so small to produce displaced vertices or even let the neutrinos escape the detector.
 
\begin{figure}
    \centering
    \begin{subfigure}[b]{0.48\textwidth}
         \centering
    \includegraphics[width=1.\textwidth]{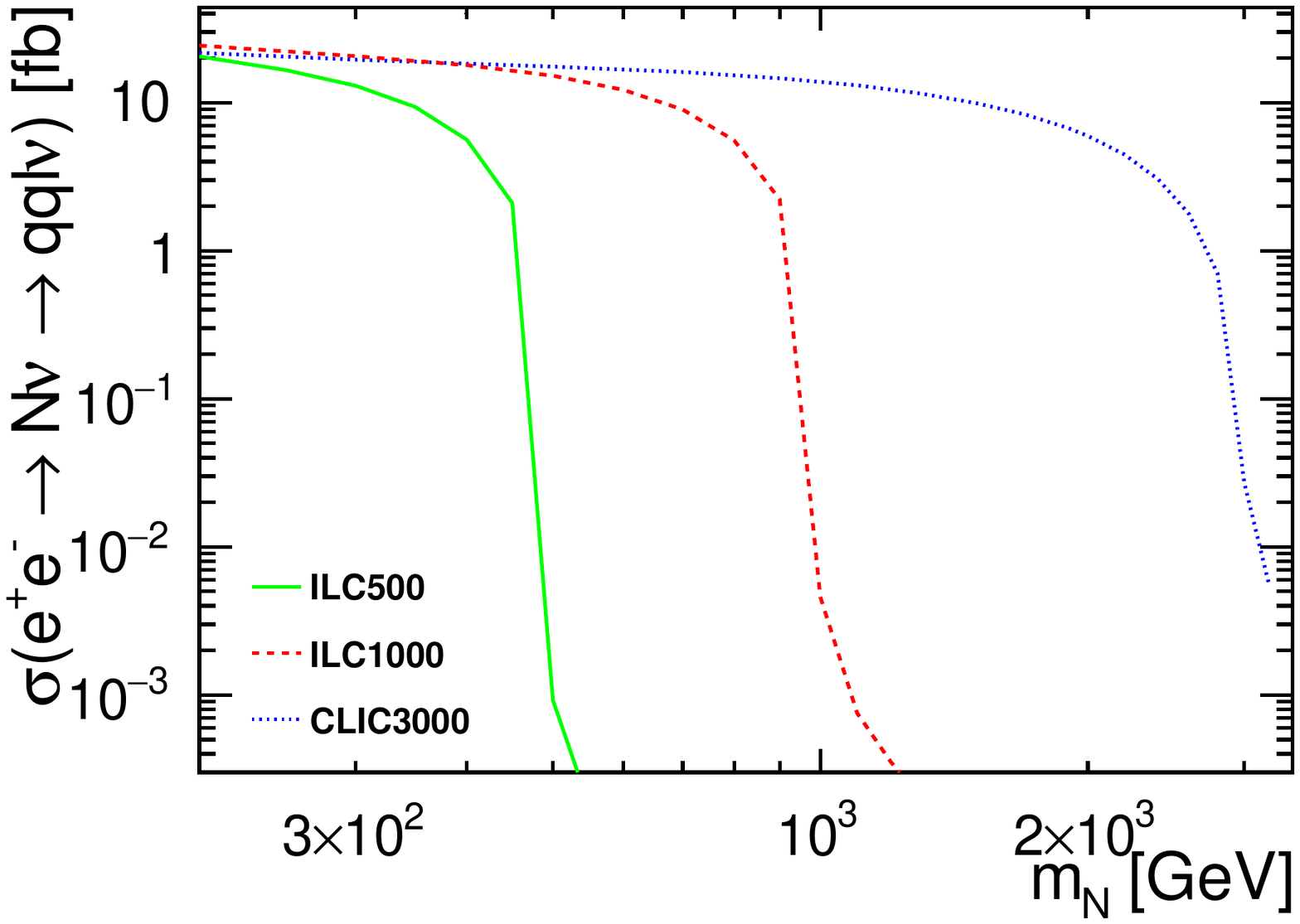}
    \caption{Dirac neutrinos}
    \end{subfigure}
    \begin{subfigure}[b]{0.48\textwidth}
         \centering
    \includegraphics[width=1.\textwidth]{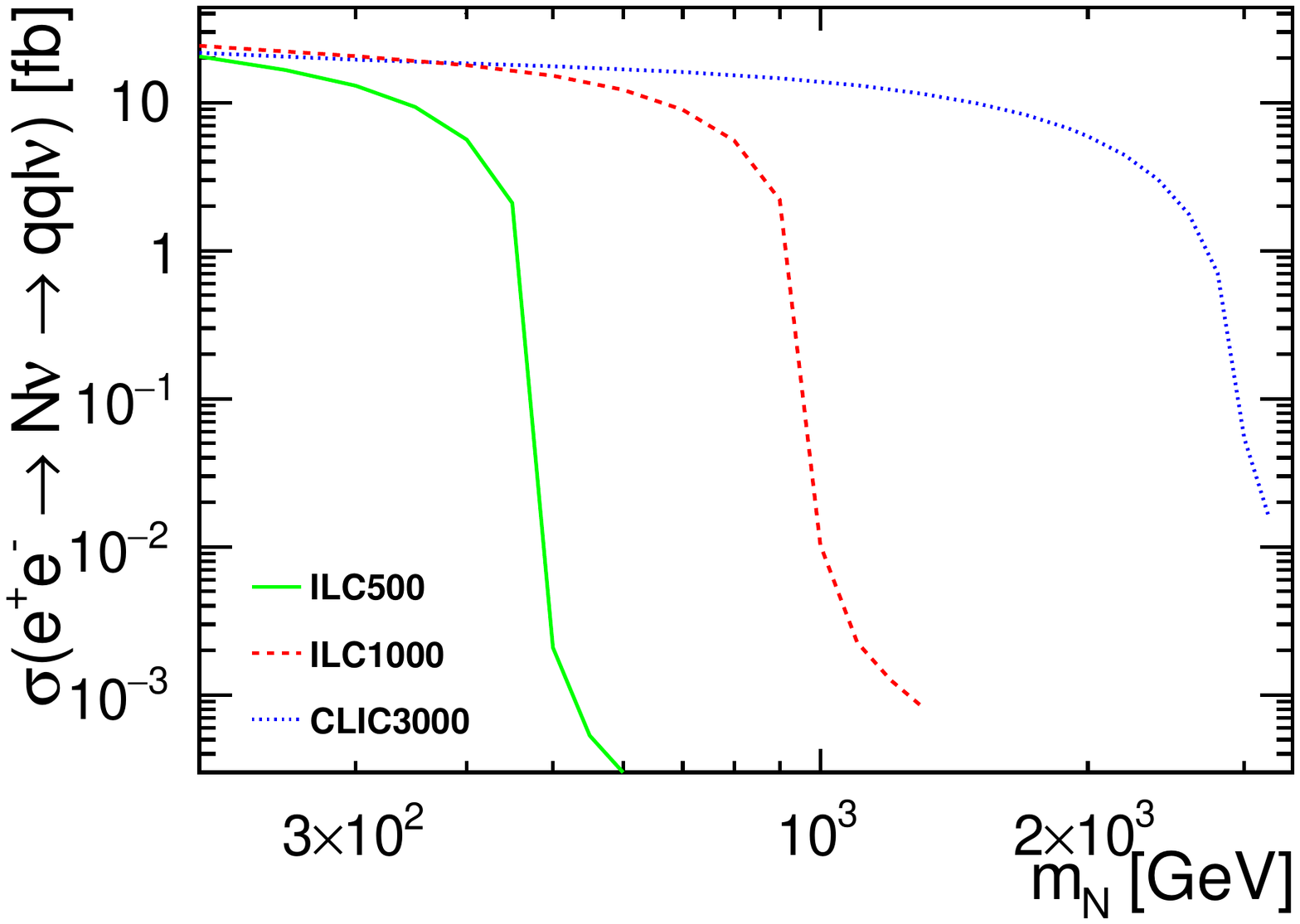}
    \caption{Majorana neutrinos}
    \end{subfigure}
    \caption{Reference scenario cross section for heavy Dirac neutrino (left plot) and Majorana neutrino (right plot) production, respectively, resulting in the $qq\ell\nu$ final state at different collider setups: ILC500 (green solid line), ILC1000 (red dashed line) and CLIC3000 (blue dotted line). Polarisation (left-right for ILC, left-unpolarised for CLIC), ISR photons and beam spectra are included. The reference scenario assumes $V_{\ell N}^{2} = 0.0003$.}
    \label{fig:cross_sections}
\end{figure}

As the signal signature, we considered the production of a light-heavy neutrino pair with the heavy neutrino decaying into two quarks (all quarks and antiquarks lighter than $t$ were allowed and their masses were set to zero in \whizard) and one lepton (all flavours allowed, only taus are assumed to be massive), so a decay $N \to \ell^\pm jj.$
For each signal scenario, 300,000 events were generated. The cross section for the process at different collider setups (including beam spectra, beam polarisation and ISR) as a function of the heavy neutrino mass is shown in Figure \ref{fig:cross_sections}. For masses below the collider energy, the cross section is of the order of 10 fb; then, it decreases fast to $10^{-2}-10^{-3}$ fb and below. It was also checked that in the wide range of couplings ($10^{-7} - 1$), the cross section can be treated as proportional to $V_{\ell N}^{2}$.

For the background samples, the Standard Model implementation (\textit{SM}) in \textsc{Whizard} was used, so the processes involving the heavy neutrino are excluded from the background. All the quark, electron and muon masses, as well as the strong coupling constant\footnote{This was assumed to avoid double counting with QCD radiation due to parton showering from \textsc{Pythia}.}, were set to zero in \whizard to assure consistency with the configuration used for the signal generation. As for the background, we considered processes with at least one lepton in the final state:
\begin{itemize}\setlength\itemsep{-0.2em}
\item $e^+e^- \to qq\ell\nu$,
\item $e^+e^- \to qq\ell\ell$,
\item $e^+e^- \to \ell\ell\ell\ell$,
\item $e^+e^- \to qq\ell\nu \ell\nu$,
\item $e^+e^- \to qqqq\ell\nu$,
\item $e^+e^- \to qqqq\ell\ell$.
\end{itemize}
Such a choice of background channels was caused by limitations of the detector simulation framework -- in \textsc{Delphes}, fake lepton tracks cannot be generated, so at least one lepton in the final state is needed. Events without any leptons would be excluded at the preselection.

Moreover, we included $\gamma$-induced background channels. Both beamstrahlung (denoted as \textit{B} in the following) and photons from collinear initial-state splittings (EPA photon interactions, denoted as \textit{E}) were considered in the analysis:
\begin{itemize}
\setlength\itemsep{-0.2em}
\item $e^+\gamma/\gamma e^-\to qq\ell$ (denoted as $\gamma e^{\pm}\to qq\ell$),
\item $\gamma\gamma\to qq\ell\nu$,
\item $\gamma\gamma\to qq\ell\ell$,
\end{itemize}
where also processes with one beamstrahlung photon and one EPA photon are taken into account.
Because of the lack of genuine \textsc{Circe2} files for the photon spectra of ILC1000, we decided on an approximate solution and scaled the ILC500 spectrum files for usage at a collision energy of 1\,TeV, as the shape of the spectra is not expected to change significantly with energy.

One should notice that the expected luminosity for the $\gamma^{B}$ collisions differs from the $e^+e^-$ luminosity. The following fractions of the integrated $e^+e^-$ luminosity are assumed in the analysis:
\begin{itemize}
\setlength\itemsep{-0.2em}
    \item ILC500: $\gamma^{B}(e^{\pm}/\gamma^{E})$ -- 57\%, $\gamma^{B}\gamma^{B}$ -- 44\%;
    \item ILC1000: $\gamma^{B}(e^{\pm}/\gamma^{E})$ -- 65\%, $\gamma^{B}\gamma^{B}$ -- 54\%;
    \item CLIC3000: $\gamma^{B}(e^{\pm}/\gamma^{E})$ -- 79\%, $\gamma^{B}\gamma^{B}$ -- 69\%.
\end{itemize}
These estimates are based on the detailed simulation of the accelerator performance~\cite{Abramowicz:2016zbo, ilcsoft}.

At the generator level, standard cuts are adopted. We require the invariant mass of the produced quark and lepton pairs to be above 10\,GeV and the four-momentum transfer between the outgoing and incoming electrons (or positrons) to be at least 4\,GeV. To avoid double-counting, for the EPA events, a maximal photon energy transfer cut of 4\,GeV is set. Furthermore, for the samples with beamstrahlung photons, we impose an additional cut on charged leptons to be detected in the central detector ($5^{\circ}<\theta<175^{\circ}$, where $\theta$ is the lepton polar angle) which helps to remove collinear singularities.

Cross sections for different processes calculated in \textsc{Whizard} are presented in Table \ref{tab:cross}.

\begin{table}
\begin{tabular}{|c||c|r||c|r||c|r|}
\hline
channel                      & $\sigma_{\textnormal{500}}$ [fb] & \multicolumn{1}{c||}{$N_{\textnormal{500}}$} & $\sigma_{\textnormal{1000}}$ [fb] & \multicolumn{1}{c||}{$N_{\textnormal{1000}}$} & $\sigma_{\textnormal{3000}}$ [fb] & \multicolumn{1}{c|}{$N_{\textnormal{3000}}$} \\ \hline
$e^+e^-\to qql\nu$ 	&	10,400	&	7,500,000	&	7,660	&	7,580,000	&	8,760	&	2,280,000	\\	
$e^+e^-\to llll$ 	&	3,010	&	60,100	&	4,190	&	73,100	&	5,810	&	838,000	\\	
$e^+e^-\to qqll$ 	&	2,020	&	158,000	&	2,510	&	188,000	&	3,210	&	204,000	\\	
$e^+e^-\to qqqqll$	&	21.9	&	5,700	&	102	&	22,900	&	701	&	5,500	\\	
$e^+e^-\to qqqql\nu$ 	&	416	&	301,000	&	209	&	265,000	&	155	&	12,800	\\	
$e^+e^-\to qql\nu l\nu$ 	&	82.7	&	52,400	&	49.0	&	53,700	&	62.6	&	22,900	\\	\hline
$\gamma^{E}e^{+}\to qql$ 	&	4,510	&	807,000	&	5,130	&	1,240,000	&	4,520	&	1,230,000	\\	
$\gamma^{E}e^{-}\to qql$ 	&	4,540	&	842,000	&	5,300	&	1,350,000	&	4,660	&	1,280,000	\\	
$\gamma^{E}\gamma^{E}\to qql\nu$ 	&	10.2	&	5,260	&	43.8	&	44,300	&	153	&	143,000	\\	
$\gamma^{E}\gamma^{E}\to qqll$ 	&	3.19	&	867	&	7.60	&	3,760	&	14.8	&	8,440	\\	\hline
$\gamma^{B}e^{+}\to qql$ 	&	8,740	&	2,550,000	&	8,290	&	4,200,000	&	4,330	&	2,110,000	\\	
$\gamma^{B}e^{-}\to qql$ 	&	8,860	&	2,610,000	&	8,660	&	4,600,000	&	4,940	&	2,530,000	\\	
$\gamma^{B}\gamma^{B}\to qql\nu$ 	&	--	&	--	&	35.6	&	35,600	&	13,500	&	13,400,000	\\	
$\gamma^{B}\gamma^{B}\to qqll$ 	&	20.4	&	5,380	&	94.1	&	64,700	&	1,460	&	1,070,000	\\	\hline
$\gamma^{E}\gamma^{B}\to qql\nu$ 	&	7.70	&	2,950	&	123	&	104,600	&	2,880	&	2,720,000	\\	
$\gamma^{E}\gamma^{B}\to qqll$ 	&	21.7	&	5,610	&	70.7	&	39,200	&	380	&	291,000	\\	\hline
\end{tabular}
\caption{
Cross section $\sigma$ and number of expected preselected events $N$ (see Section \ref{sec:analysis}) for different channels at ILC500, ILC1000 and CLIC3000. The cross section for $\gamma^{B}\gamma^{B}\to qql\nu$ at ILC500 is negligible (0.042 fb) because the energy spectrum of the photons is too low for the on-shell $WW$ production.}
\label{tab:cross}
\end{table}

\subsection{Detector simulation}
In the next step, the fast detector simulation framework \textsc{Delphes}~\cite{deFavereau:2013fsa} was used to simulate the detector response, with cards available for parameterisation of the ILC detector (\textit{delphes\_card\_ILCgen.tcl}) and CLIC detector (\textit{delphes\_card\_CLICdet\_Stage3\_fcal.tcl}), respectively. As opposed to programs based on \textit{full simulation}, \textsc{Delphes} provides a general parametrisation of the detector acceptance and response, making the simulation much faster than in the standard approach and allowing for testing many points in the parameter space. 
In the ILC detector model, the Durham algorithm was implemented for jet reconstruction, following results of the full simulation studies \cite{ILD:2020qve}, while for CLIC, the VLC algorithm with the following parameter setup: $R$ = 0.5, $\beta$ = 1, $\gamma$ = 1 (see \cite{Boronat:2016tgd} for details) was applied. Results of the clustering in the exclusive two-jet mode were selected for the presented study based on the expected signal topology.

\section{Analysis procedure}
\label{sec:analysis}
The first step of the analysis was to exclude events resulting in a different topology than the one expected for the signal. 
Only events consisting of two jets and one lepton (electron or muon) were accepted. Events with any other activity in the detector (additional leptons or reconstructed photons) were rejected. 
It was also required that the total transverse momentum of final state objects not contributing to the required final state (untagged transverse momentum) had to be smaller than 20 GeV.
In particular, this cut rejects events with significant contribution of forward deposits assigned to the beam jets (not included in the final state) by the VLC algorithm.
One should notice that events with the $qq\tau\nu$ final state could also pass the preselection, if the $\tau$ decayed into leptons. Numbers of expected events passing the above cuts at the considered future collider options are given in Table \ref{tab:cross}.

\begin{figure}
    \centering
    \begin{subfigure}[b]{0.48\textwidth}
         \centering
         \includegraphics[width=\textwidth]{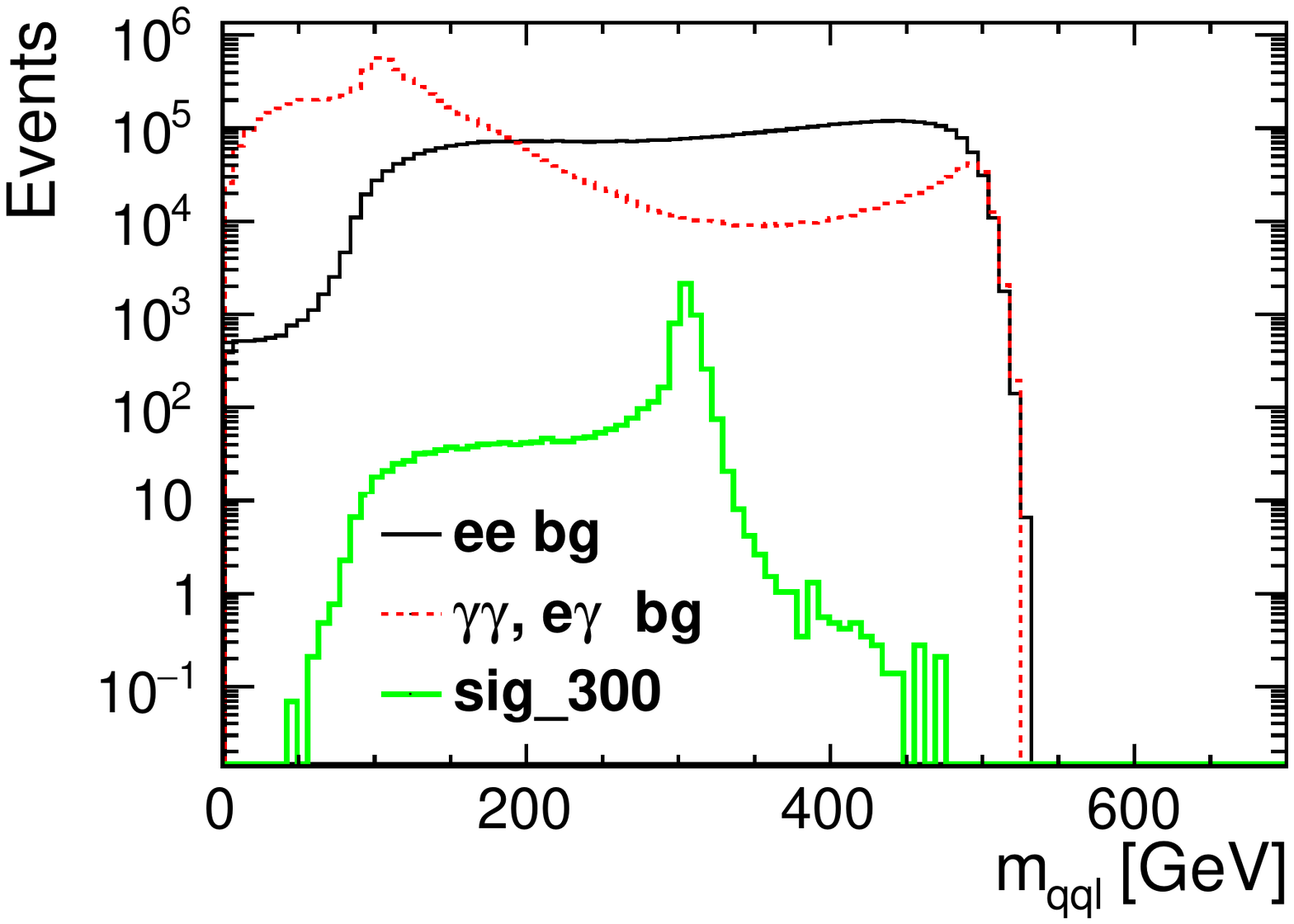}
         \caption{ILC500 with electrons}
     \end{subfigure}
    \begin{subfigure}[b]{0.48\textwidth}
         \centering
         \includegraphics[width=\textwidth]{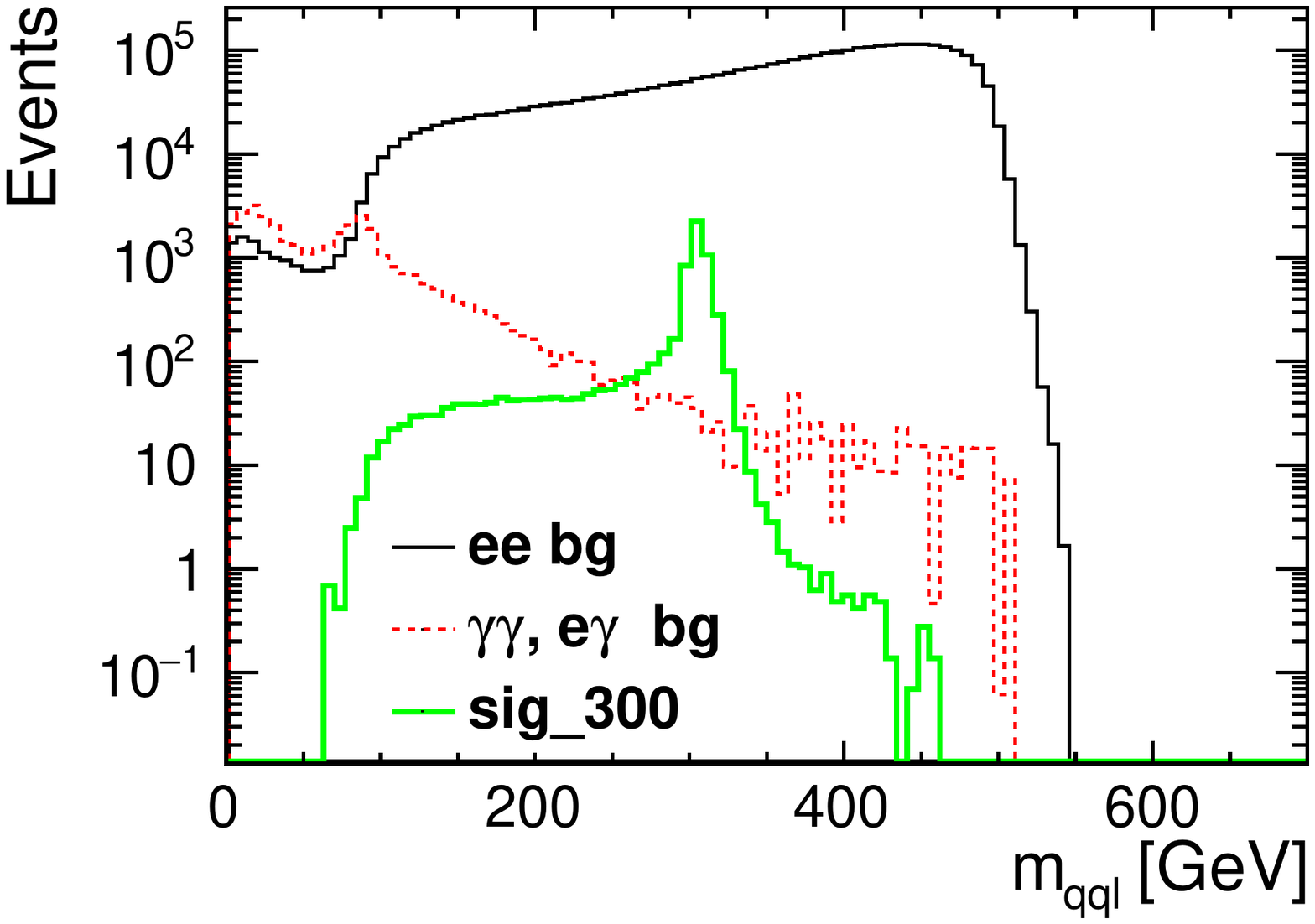}
         \caption{ILC500 with muons}
     \end{subfigure}
     \begin{subfigure}[b]{0.48\textwidth}
         \centering
         \includegraphics[width=\textwidth]{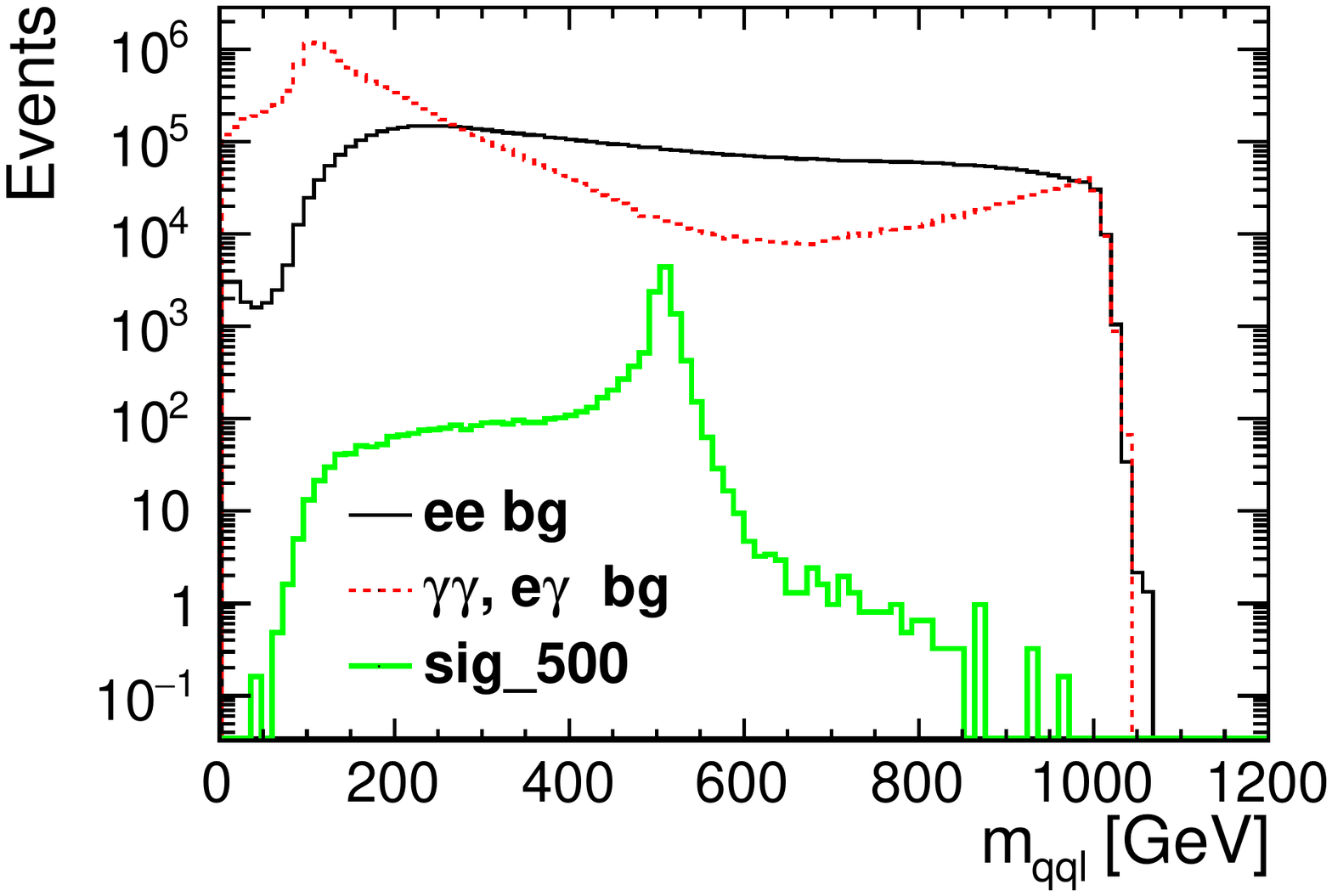}
         \caption{ILC1000 with electrons}
     \end{subfigure}
     \begin{subfigure}[b]{0.48\textwidth}
         \centering
         \includegraphics[width=\textwidth]{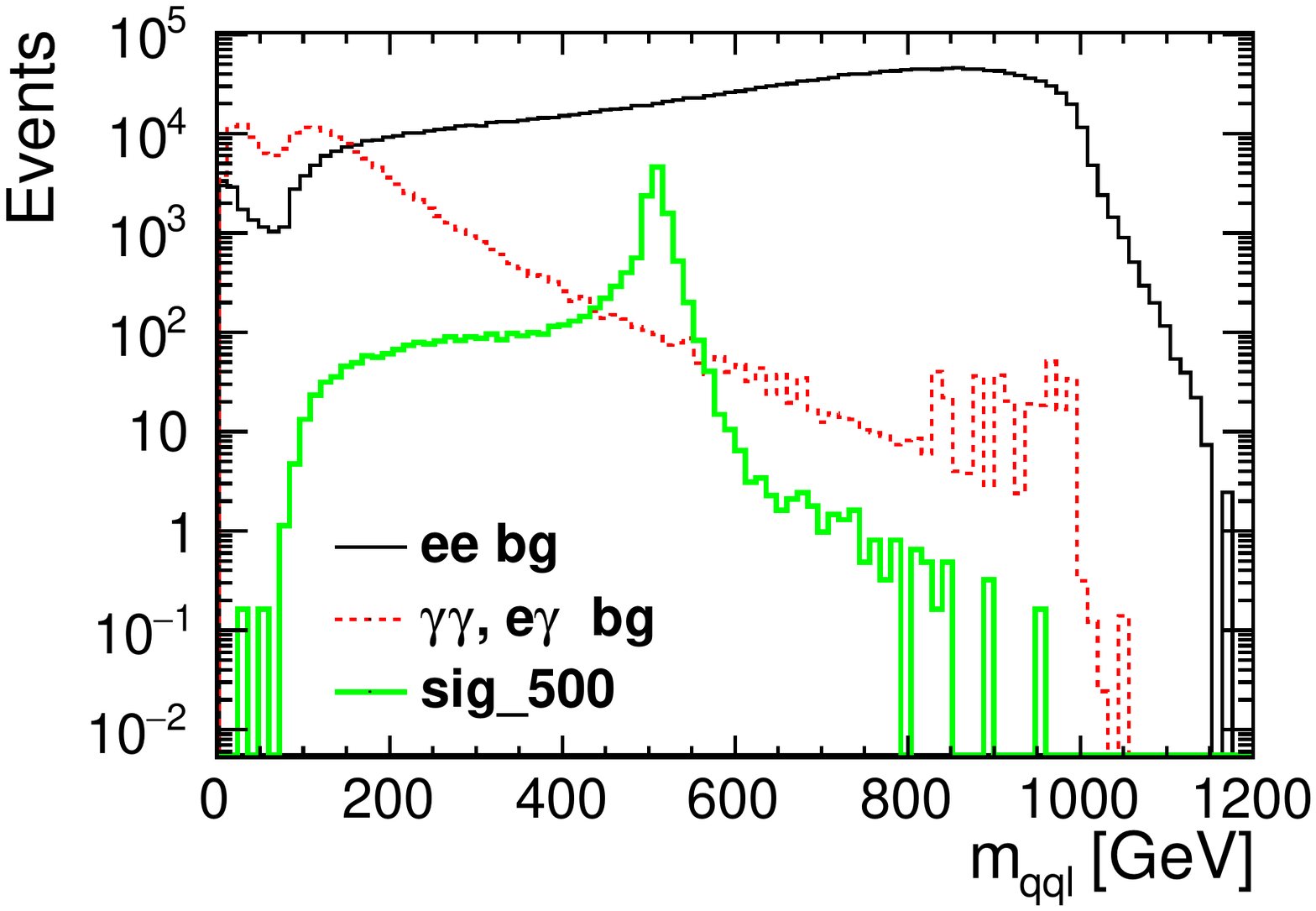}
         \caption{ILC1000 with muons}
     \end{subfigure}
     \begin{subfigure}[b]{0.48\textwidth}
         \centering
         \includegraphics[width=\textwidth]{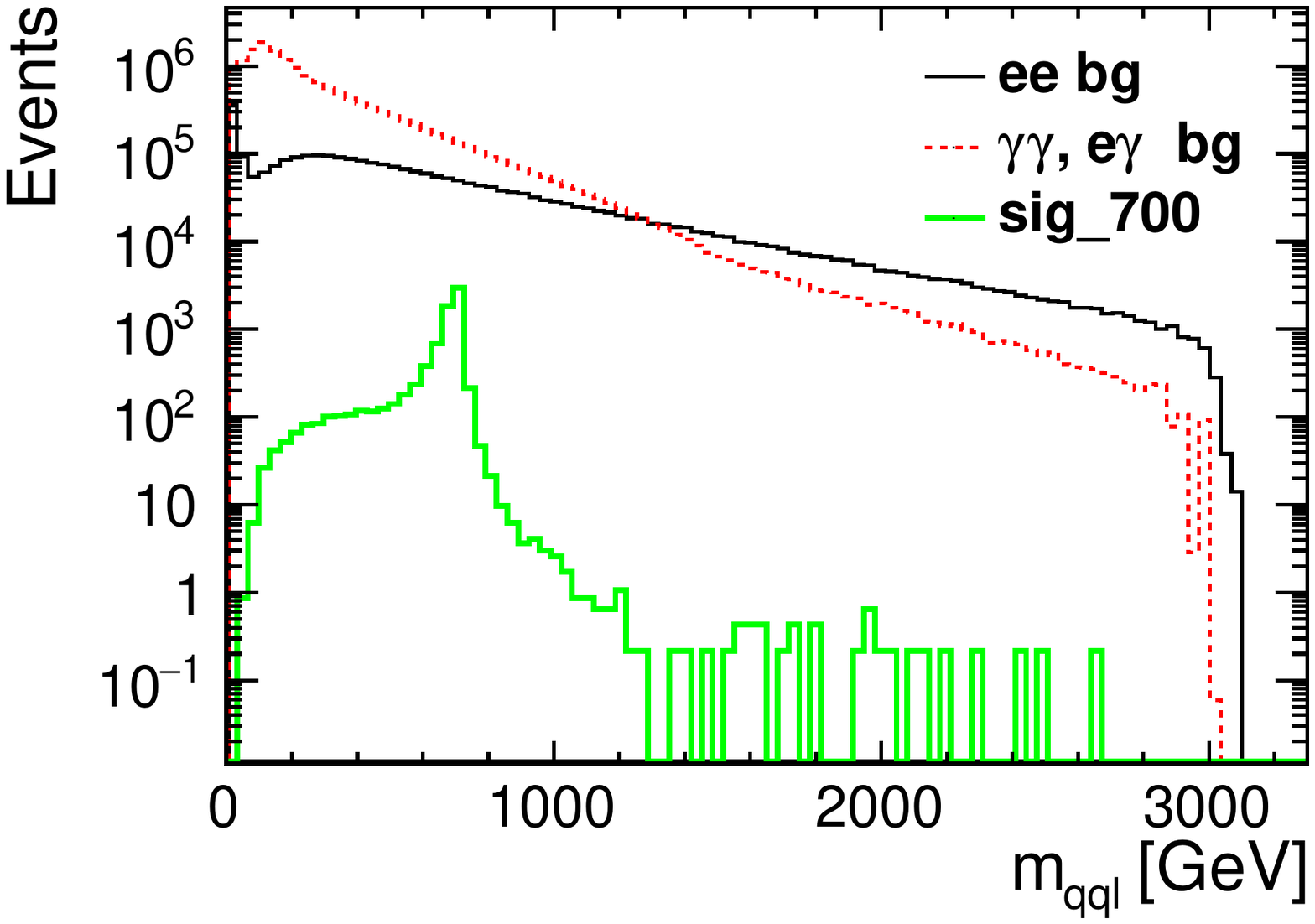}
         \caption{CLIC3000 with electrons}
     \end{subfigure}
     \begin{subfigure}[b]{0.48\textwidth}
         \centering
         \includegraphics[width=\textwidth]{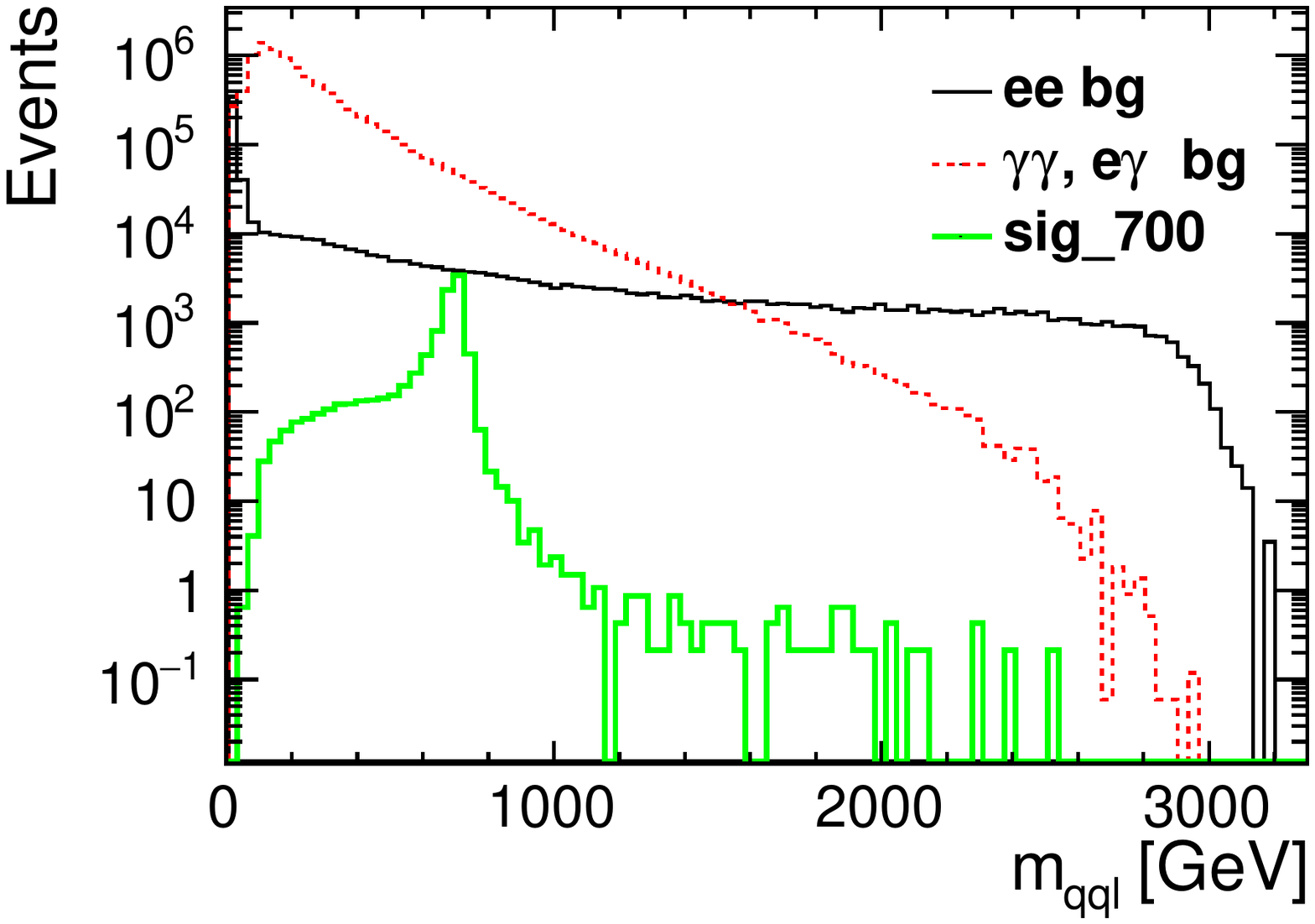}
         \caption{CLIC3000 with muons}
     \end{subfigure}
    \caption{$qq\ell$ mass distribution for different collider setups for electrons (left) and muons (right) in the final state, respectively. Black solid lines stand for the $e^+e^-$ background, red dashed lines for the $\gamma$-induced background and thick green lines for the different signal scenarios -- results for Dirac neutrinos with masses of 300\,GeV, 500\,GeV and 700\,GeV are presented for ILC500, ILC1000 and CLIC3000, respectively.}
    \label{fig:mass}
\end{figure}

In Figure \ref{fig:mass}, distributions of the invariant mass of two jets and a lepton are shown for different collider setups. A clear peak corresponding to the heavy neutrino mass is visible in each plot. The left shoulders of those peaks can be explained by the contribution of leptonic $\tau$ decays, when two additional escaping neutrinos reduce the invariant mass of the detectable final state. The tails on the right-hand side are caused by detector effects, for example, worse track momentum resolution for leptons going at small angles. It is also important to notice that the background levels for the muon channel are significantly smaller. An extra cut on the invariant mass could help with the background-signal separation at the preselection level, but we do not apply it, as we want to consider broad spectra of heavy neutrino mass values. Nevertheless, it was checked that the cut does not affect the final results obtained with the Boosted Decision Tree algorithm.

In the next step, the Boosted Decision Tree (BDT) method implemented in the \textit{TMVA} package~\cite{Hocker:2007ht} was used to discriminate between signal and background events.
The following 8 variables were considered to train the BDT algorithm:
\begin{itemize}\setlength\itemsep{-0.2em}
\item m$_{qq\ell}$ -- invariant mass of the dijet-lepton system,
\item $\alpha$ -- angle between the dijet-system and the lepton,
\item $\alpha_{qq}$ -- angle between the two jets,
\item E$_{\ell}$ -- lepton energy,
\item E$_{qq\ell}$ -- energy of the dijet-lepton system,
\item p$^{T}_{\ell}$ -- lepton transverse momentum,
\item p$^{T}_{qq}$ -- dijet transverse momentum,
\item p$^{T}_{qq\ell}$ -- transverse momentum of the dijet-lepton system.
\end{itemize}
Other variables were also investigated, but it was found that they did not improve the BDT performance. 

The BDT algorithm was trained separately for events with electrons and muons in the final state. The main reason for this approach was the fact that there are more background channels for electrons in the final state and the results for this case were expected to be less stringent.

\begin{figure}
    \centering
    \includegraphics[width=0.8\textwidth]{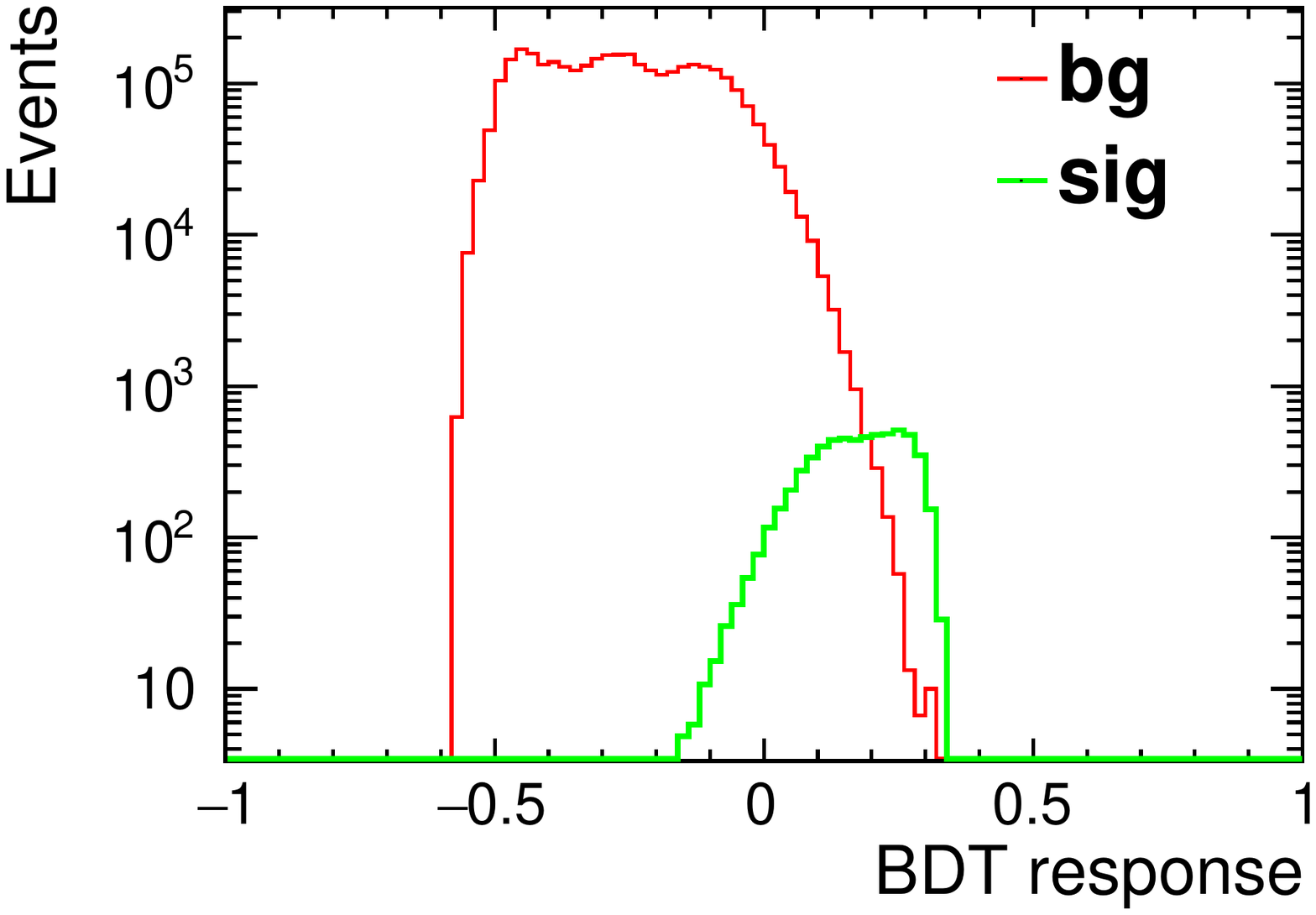}
    \caption{Distribution of the BDT response for the reference scenario (Dirac neutrino, m$_N$ = 300\,GeV) with muons at ILC500. The red line denotes the background, the green line the signal.}
    \label{fig:BDT_response}
\end{figure}

\begin{figure}
    \centering
    \begin{subfigure}[b]{0.45\textwidth}
         \centering
         \includegraphics[width=\textwidth]{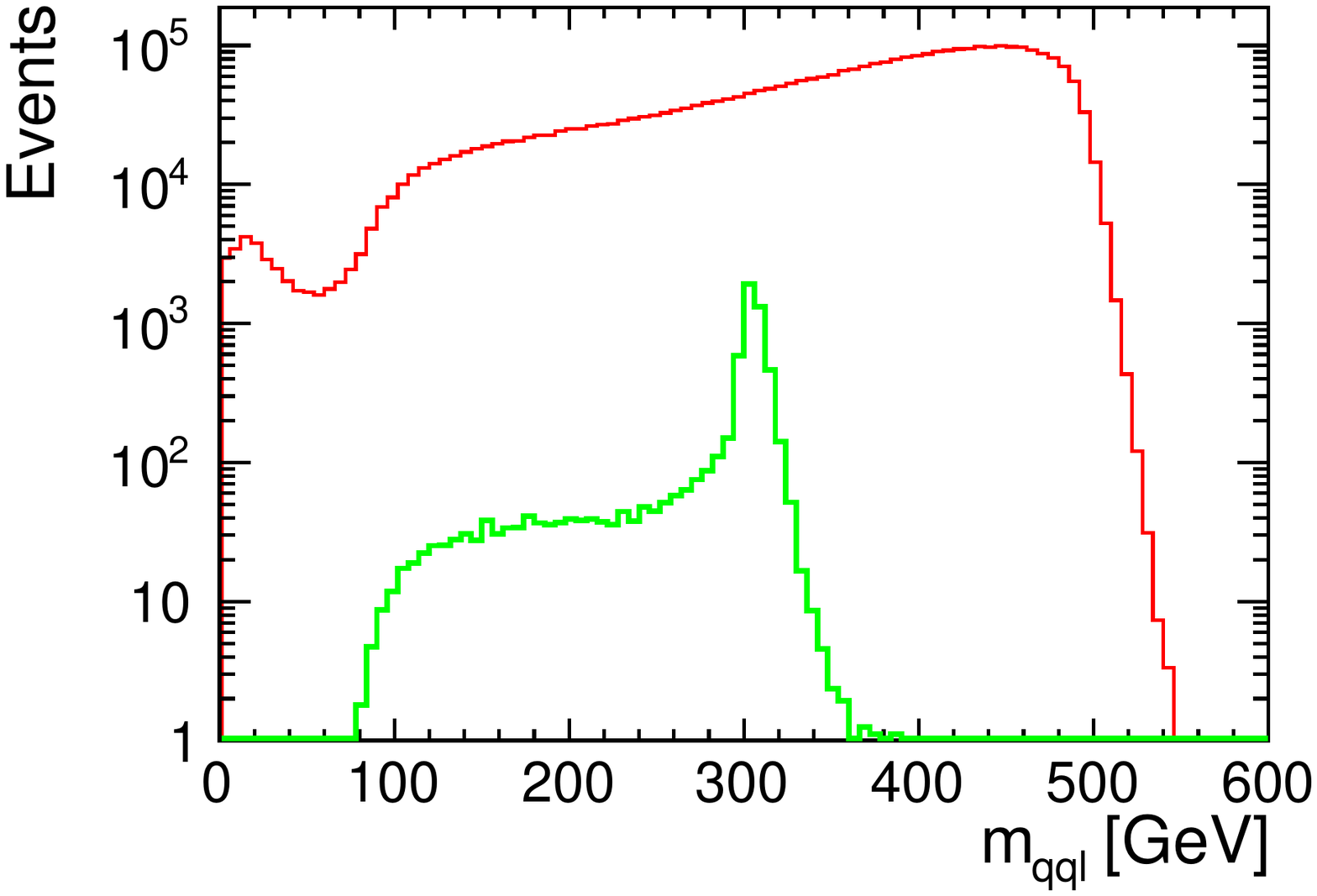}\\[-5mm]
         \caption{m$_{qq\ell}$}
    \end{subfigure}
    \begin{subfigure}[b]{0.45\textwidth}
         \centering
         \includegraphics[width=\textwidth]{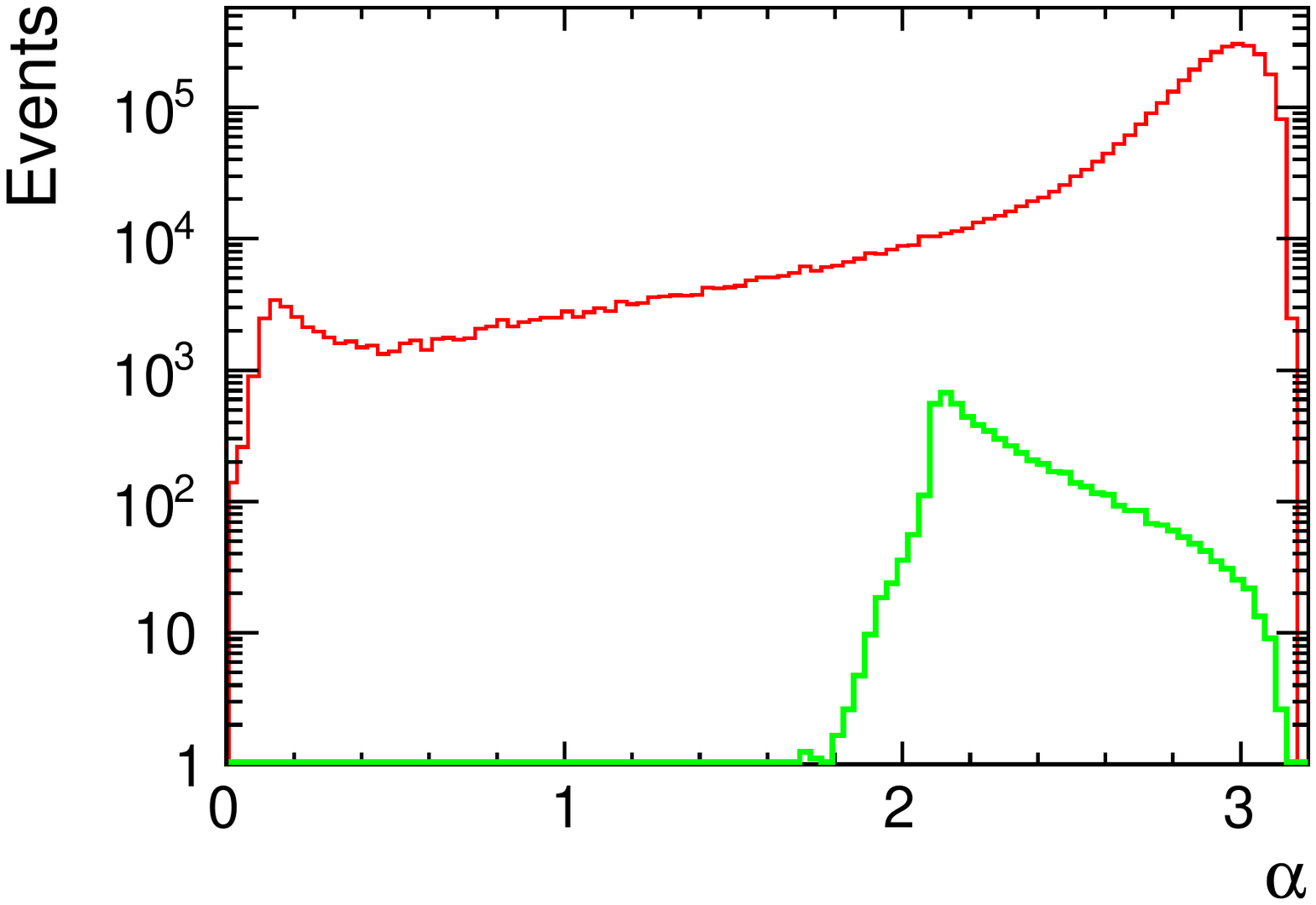}\\[-5mm]
         \caption{$\alpha$}
    \end{subfigure}
    \begin{subfigure}[b]{0.45\textwidth}
         \centering
         \includegraphics[width=\textwidth]{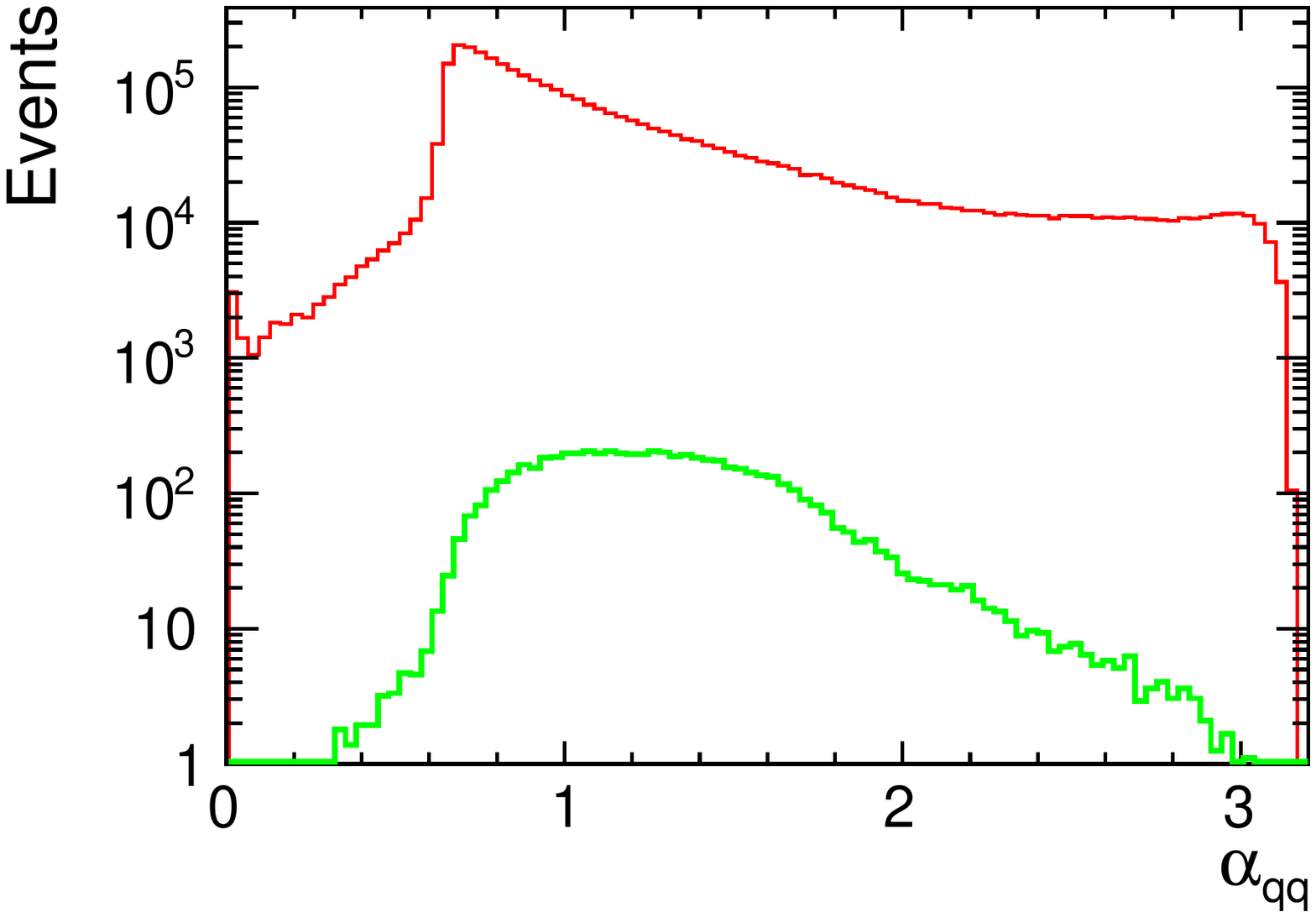}\\[-5mm]
         \caption{$\alpha_{qq}$}
    \end{subfigure}
    \begin{subfigure}[b]{0.45\textwidth}
         \centering
         \includegraphics[width=\textwidth]{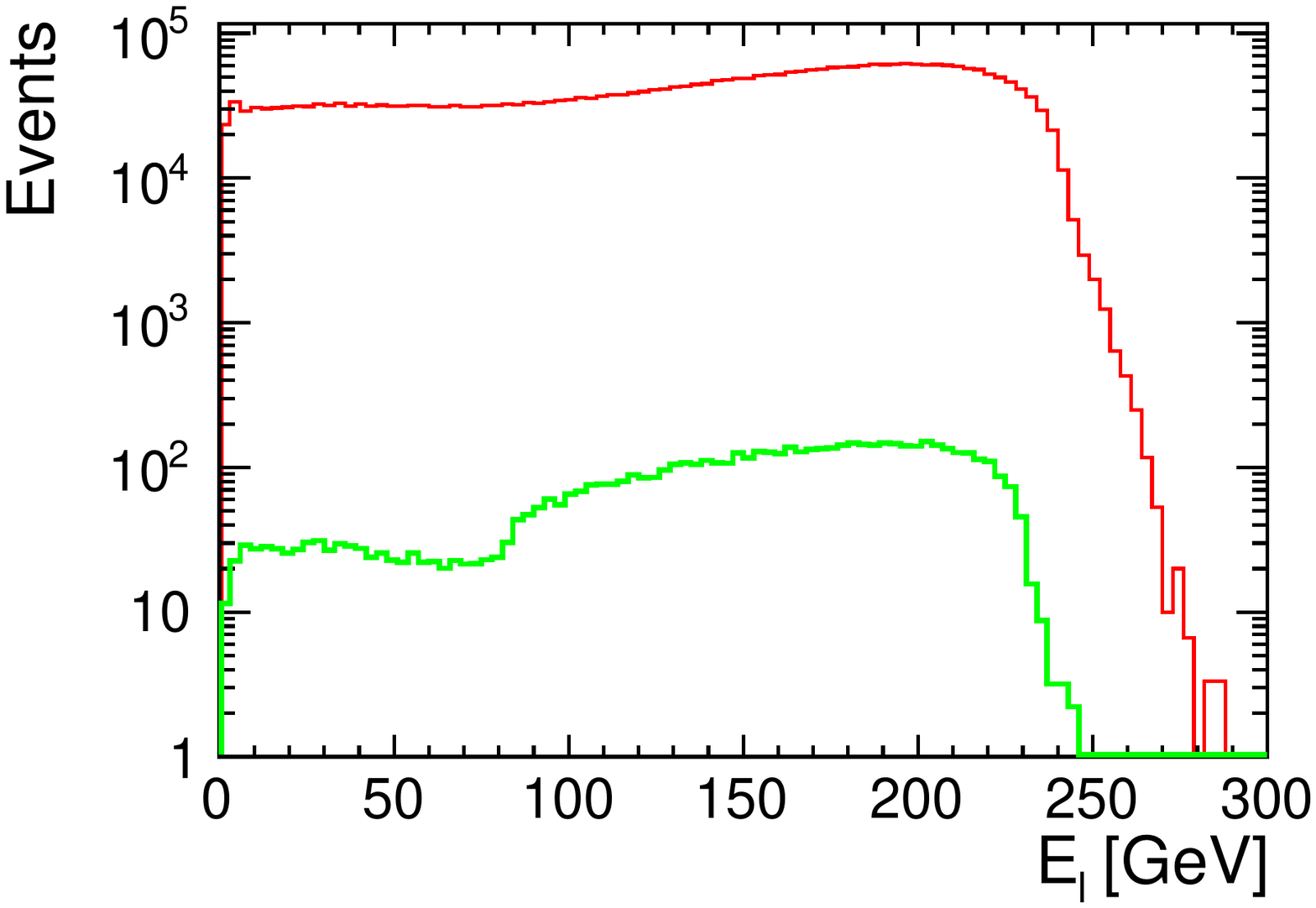}\\[-5mm]
         \caption{E$_{\ell}$}
    \end{subfigure}
    \begin{subfigure}[b]{0.45\textwidth}
         \centering
         \includegraphics[width=\textwidth]{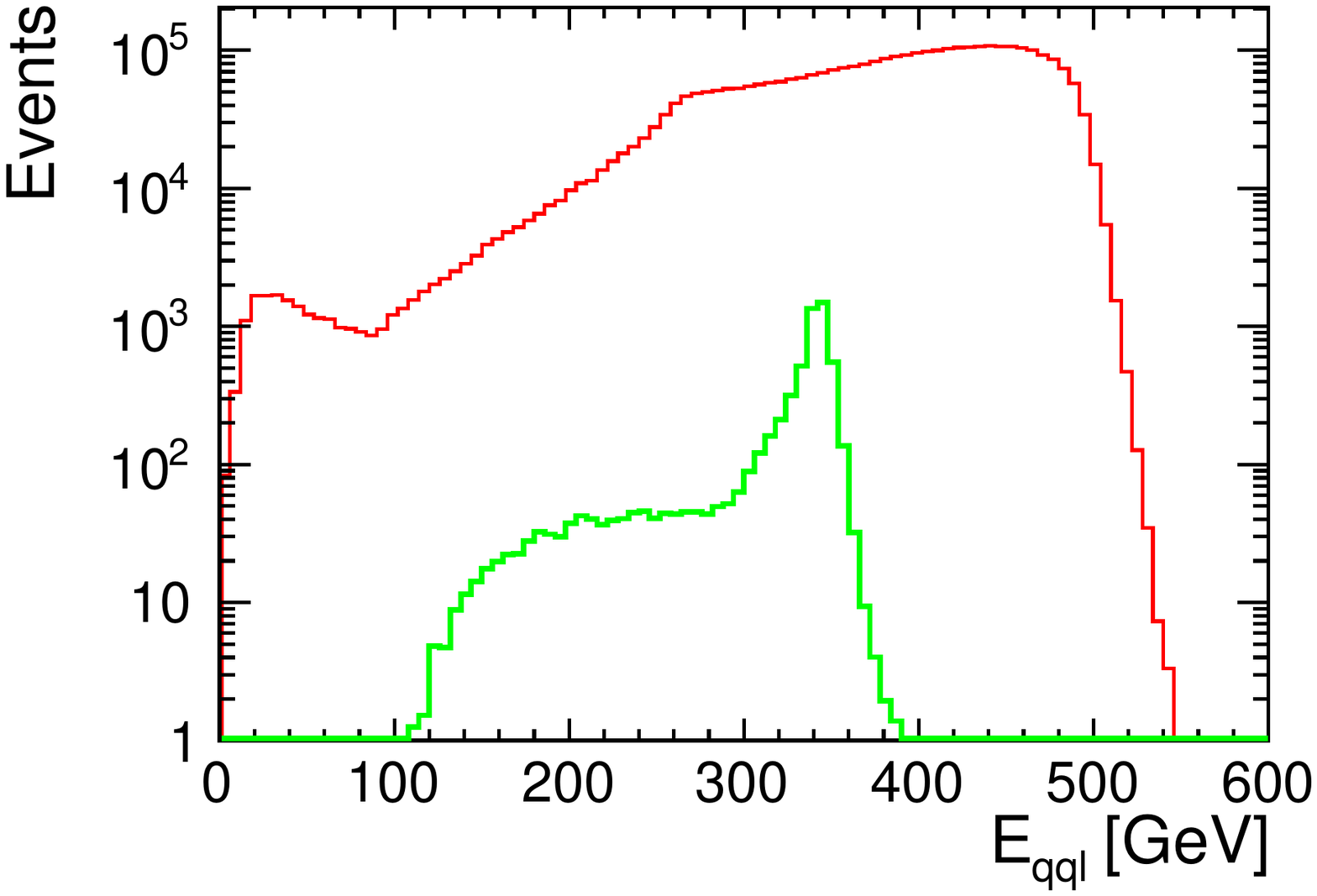}\\[-5mm]
         \caption{E$_{qq\ell}$}
    \end{subfigure}
    \begin{subfigure}[b]{0.45\textwidth}
         \centering
         \includegraphics[width=\textwidth]{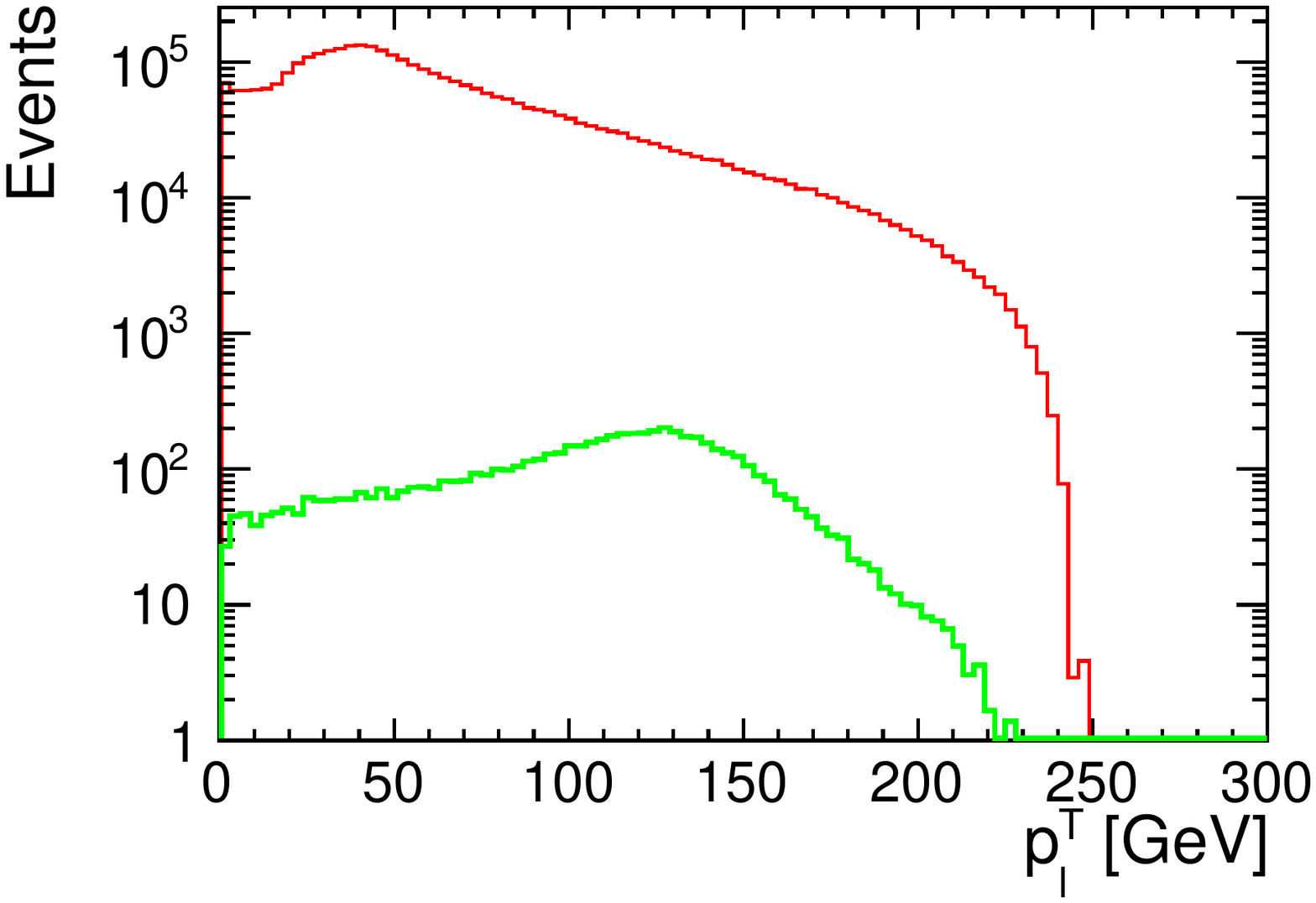}\\[-5mm]
         \caption{p$^{T}_{\ell}$}
    \end{subfigure}
    \begin{subfigure}[b]{0.45\textwidth}
         \centering
         \includegraphics[width=\textwidth]{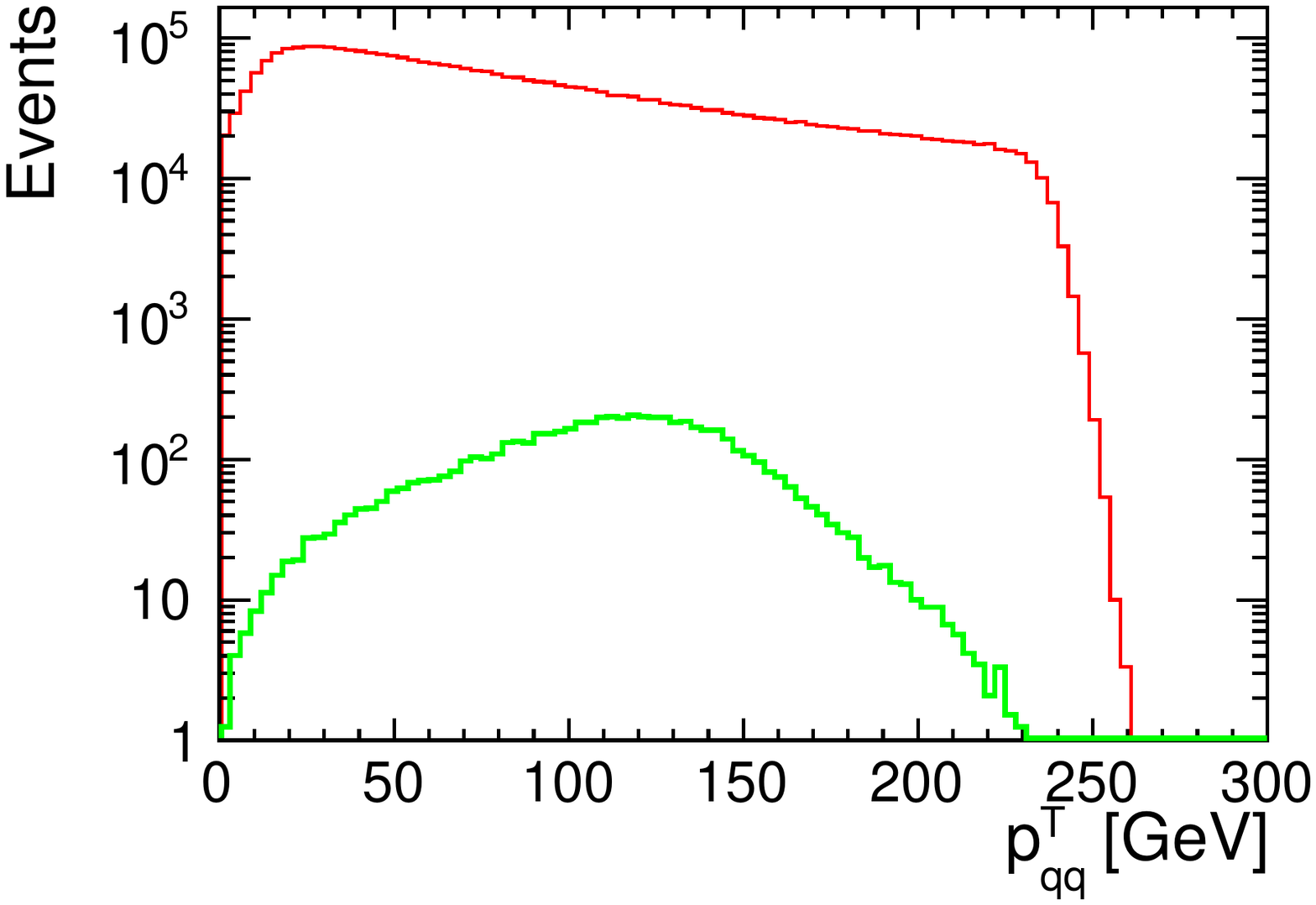}\\[-5mm]
         \caption{p$^{T}_{qq}$}
    \end{subfigure}
    \begin{subfigure}[b]{0.45\textwidth}
         \centering
         \includegraphics[width=\textwidth]{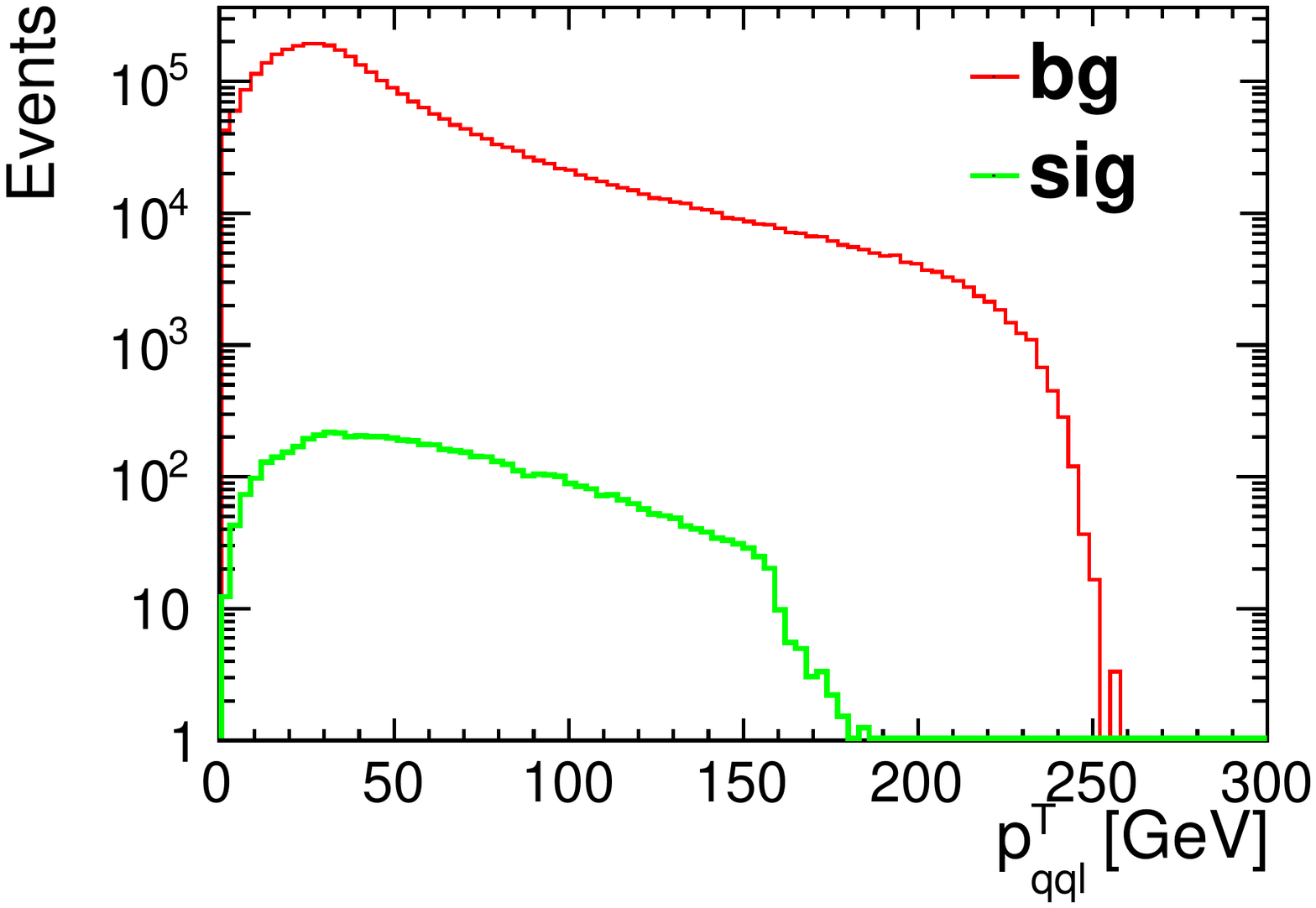}\\[-5mm]
         \caption{p$^{T}_{qq\ell}$}
    \end{subfigure}
    \caption{Distributions of the variables used in the BDT procedure for the reference scenario (Dirac neutrino, m$_N$ = 300\,GeV) with muons at ILC500. The red line stands for the background, the green line for the signal.}
    \label{fig:BDT_variables}
\end{figure}

The BDT response for an example reference scenario (Dirac neutrino, m$_N$ = 300\,GeV) with muons in the final state at ILC500 is shown in Figure \ref{fig:BDT_response}. In Figure \ref{fig:BDT_variables}, the variable distributions for the same scenario are presented.

In the last step, the CL$_s$ method, implemented within the \textit{RooStats} package~\cite{Moneta:2010pm}, was used to extract the cross section limits from the expected BDT response distributions.
This approach allows for combining different measurement channels (electrons and muons in this case) and adding systematic uncertainties. We considered only systematics related to the normalisation of the samples. The normalisation uncertainty of the $e^+e^-$ data sets was set to 1\%, and for the $\gamma^{BS}e^{\pm}$ and $\gamma^{BS}\gamma^{BS}$ backgrounds additional uncertainties of 2\% and 5\%, respectively, were applied. These values can be treated as conservative (see e.g. \cite{Habermehl:2020njb}), but it was verified that even without a normalisation constraint (i.e., setting the normalisation uncertainty to 100\%), the extracted limits are hardly changed.

We also verified the effect of the jet energy scale uncertainty for a few example points in the parameter space. Jet energy-momentum 4-vectors were scaled up and down by 1\%. Since it turned out that there is no impact on the final results, we refrained from studying the effect. Other kinds of uncertainties are also not expected to affect the final conclusions significantly and thus, were not included in the analysis procedure.

\section{Results}
\begin{figure}
    \centering
    \begin{subfigure}[b]{0.47\textwidth}
         \centering
         \includegraphics[width=\textwidth]{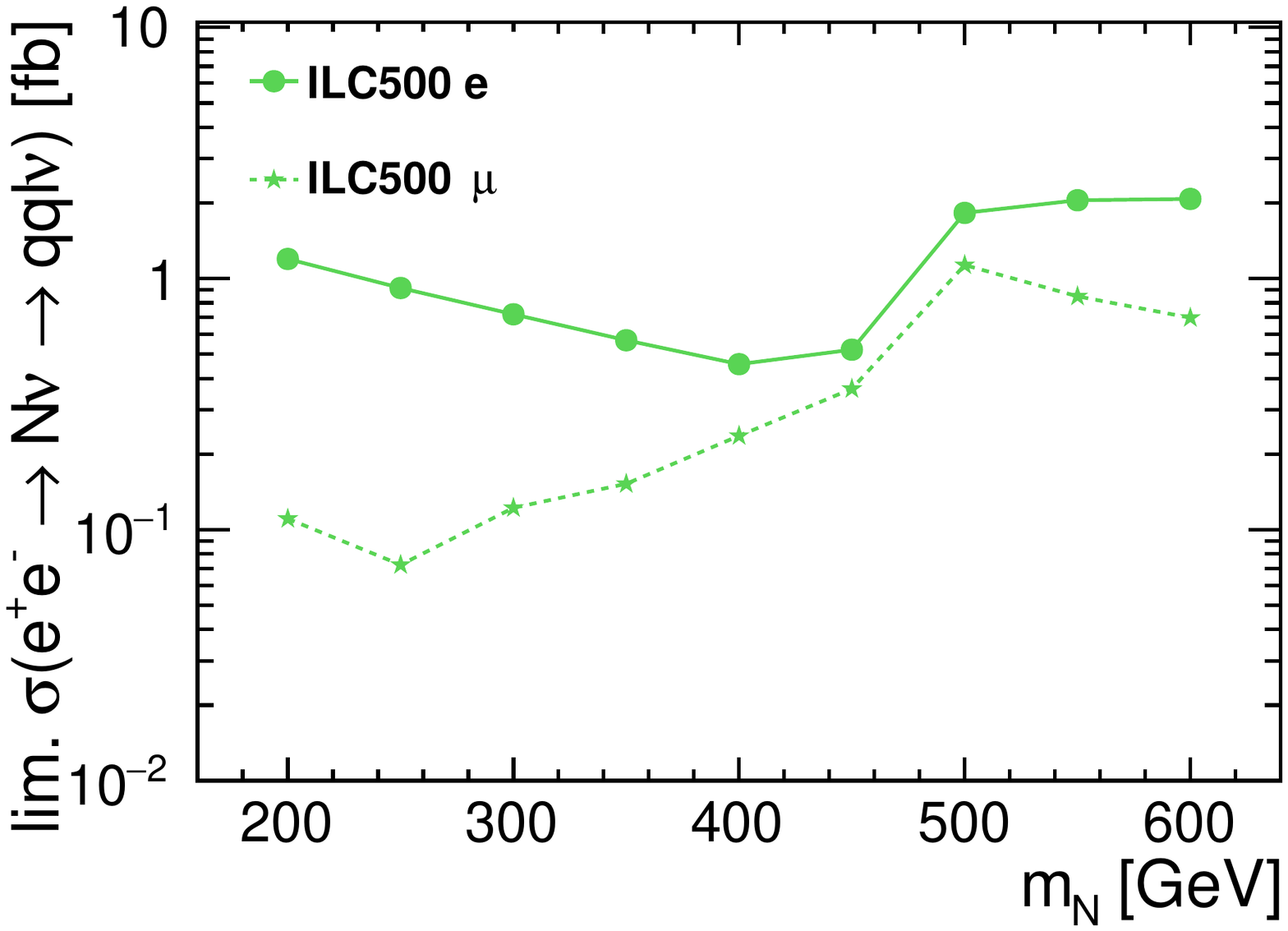}
         \caption{ILC500}
     \end{subfigure}
     \begin{subfigure}[b]{0.47\textwidth}
         \centering
         \includegraphics[width=\textwidth]{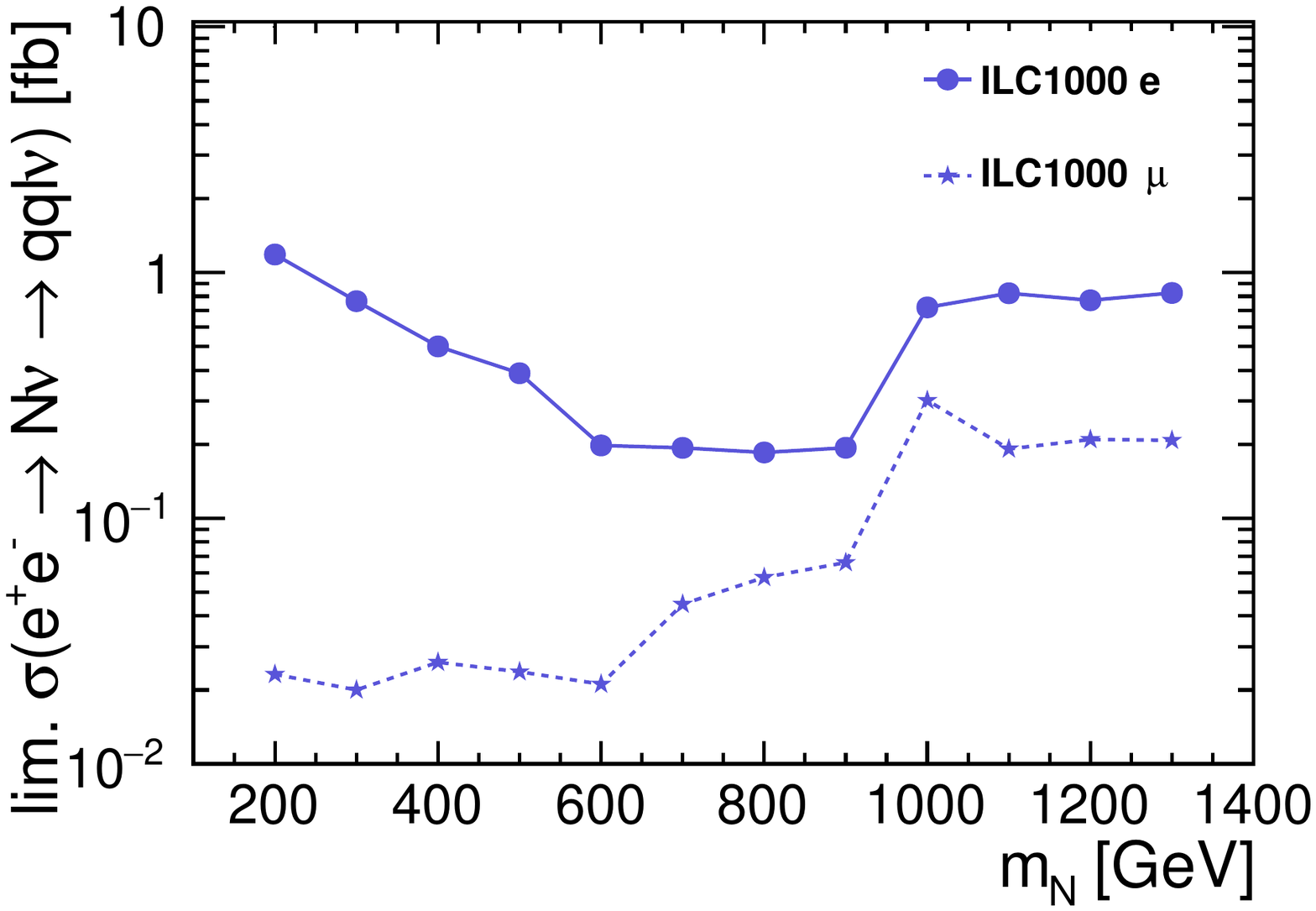}
         \caption{ILC1000}
     \end{subfigure}
     \begin{subfigure}[b]{0.47\textwidth}
         \centering
         \includegraphics[width=\textwidth]{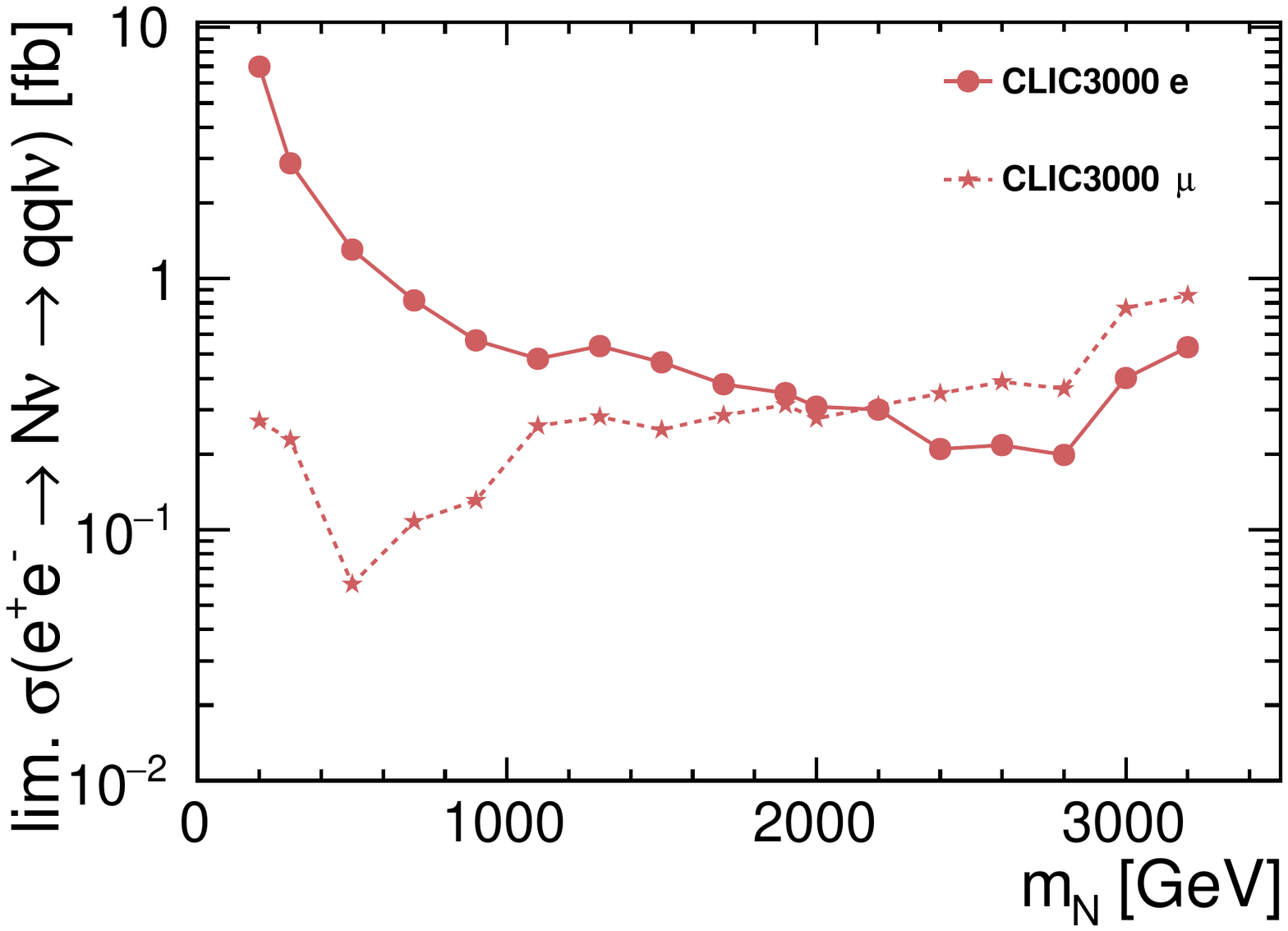}
         \caption{CLIC3000}
     \end{subfigure}
    \caption{95\% C.L. limits on the cross section of heavy Dirac neutrino production and decay (the $qq\ell\nu$ final state) as a function of the neutrino mass for different collider setups. Dots signify the analysis with an electron in the final state and stars that with a muon, respectively.}
    \label{fig:cross_section_limits}
\end{figure}

After having detailed the analysis methods, we present in this section our results for the sensitivity of ILC and CLIC to heavy neutrinos. In Figure \ref{fig:cross_section_limits}, the limits on the cross section for the considered process are presented, separately for electron and muon channel studies. Better limits for most of the considered scenarios are obtained for muons. Only for the highest neutrino masses at CLIC3000, the limits resulting from the electron channel are slightly stronger.

\begin{figure}[h]
    \centering
    \includegraphics[width=0.8\textwidth]{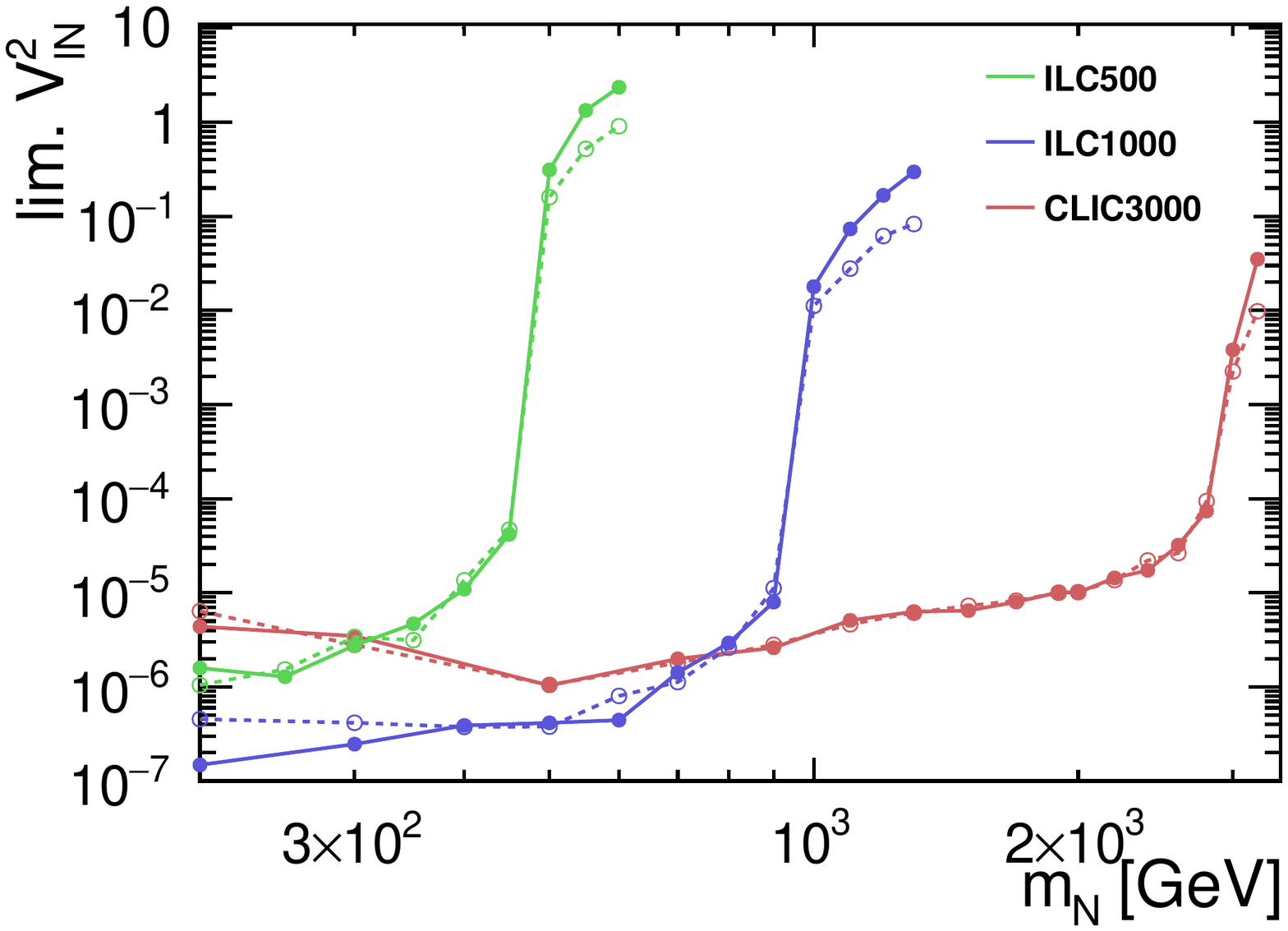}
    \caption{Comparison between results for Majorana (dashed line) and Dirac (solid line) neutrinos for different collider scenarios: green for ILC500, blue for ILC1000 and red for CLIC3000 respectively.}
    \label{fig:Majorana}
\end{figure}

Extracted cross section limits result from the expected background level and the signal selection efficiency after the optimised event selection. They only reflect the experimental sensitivity and do not depend on the signal cross section predictions.
This is why the cross section limits do not get significantly weaker for neutrino masses above the collision energy. Processes mediated by off-shell neutrino exchange are also included in our analysis and signal-background discrimination is only slightly weaker in this case. However, the cross sections for such processes are much smaller than for the on-shell production (refer to Fig.~\ref{fig:cross_sections}), so the corresponding limits on the neutrino coupling $V_{\ell N}^{2}$ are much weaker.
Such limits are presented in Figure \ref{fig:Majorana}, where combined results for Dirac and Majorana neutrino hypotheses are compared. Limits for the two neutrino types are very similar in a wide range of neutrino masses. Below the energy threshold, the differences could be interpreted as statistical fluctuations. However, above the threshold, a separation between the lines is clearly visible. The reason for such a behaviour is the fact that for large neutrino masses, off-shell production above the collider energy is more sensitive to the neutrino width. Since the width of the heavy Dirac neutrino is larger by a factor of 2, so is the production cross section (see Figure \ref{fig:cross_sections}), and more events are expected to be observed for the same coupling value, resulting in stronger limits.

\begin{figure}[t]
    \centering
    \includegraphics[width=0.8\textwidth]{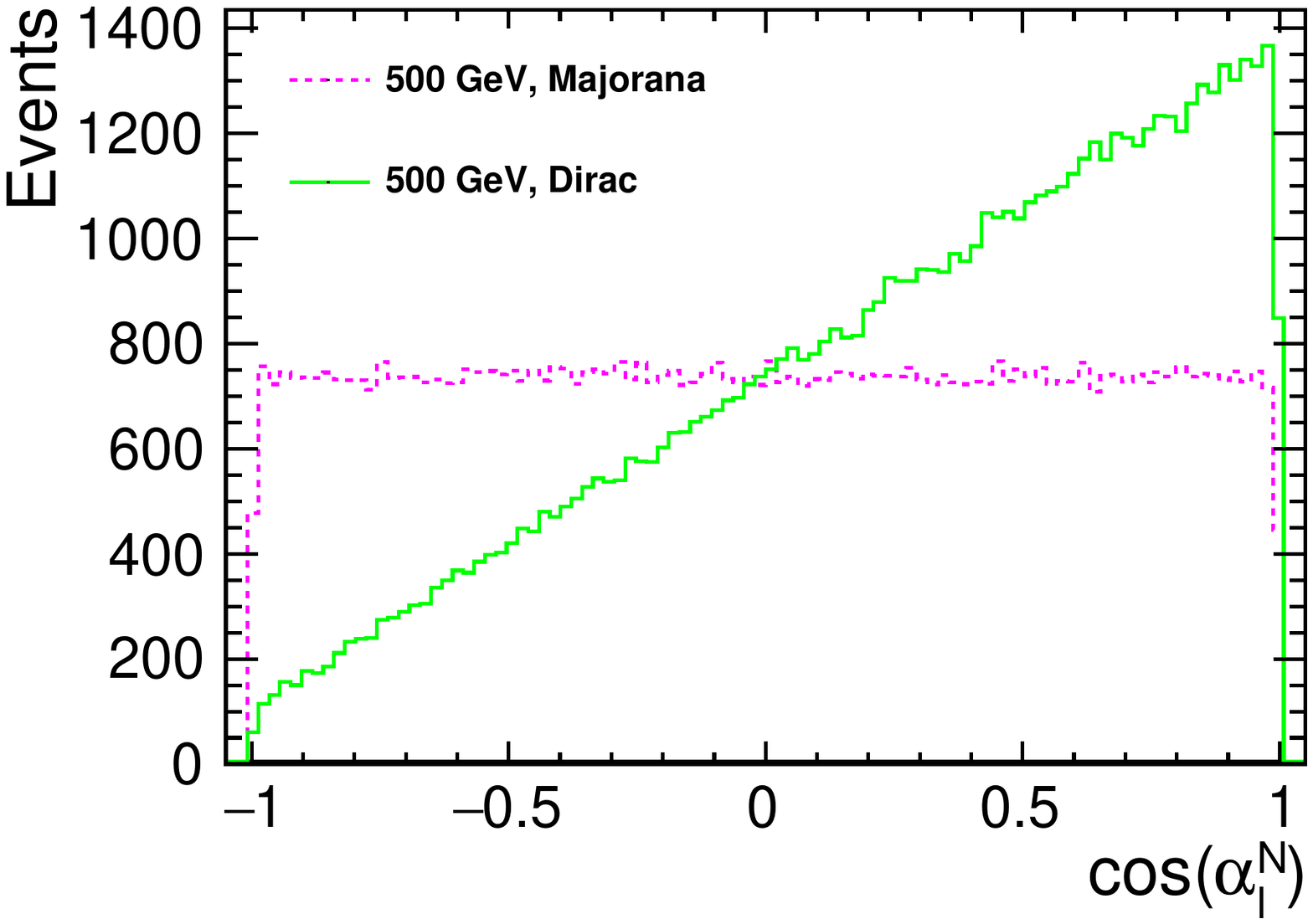}
    \caption{Distribution of the cosine of the lepton emission angle in the $N$ rest-frame for the Majorana (pink dashed line) and the Dirac (green solid line) neutrinos with a mass of 500\,GeV at CLIC3000 (generator level)}
    \label{fig:DiMaj_angle}
\end{figure}

\begin{figure}
    \centering
     \begin{subfigure}[b]{0.49\textwidth}
         \centering
         \includegraphics[width=\textwidth]{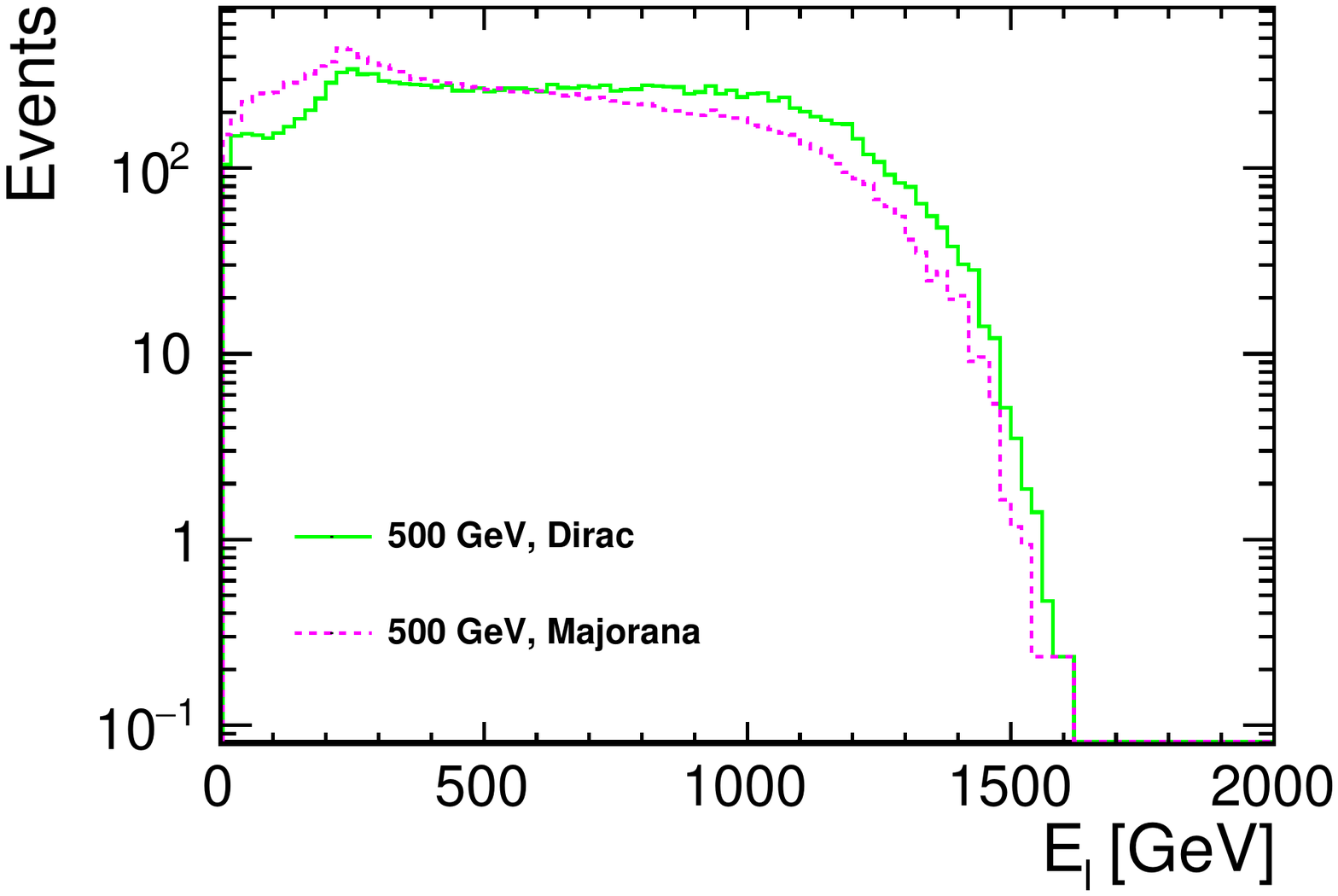}
     \end{subfigure}
    \begin{subfigure}[b]{0.49\textwidth}
         \centering
         \includegraphics[width=\textwidth]{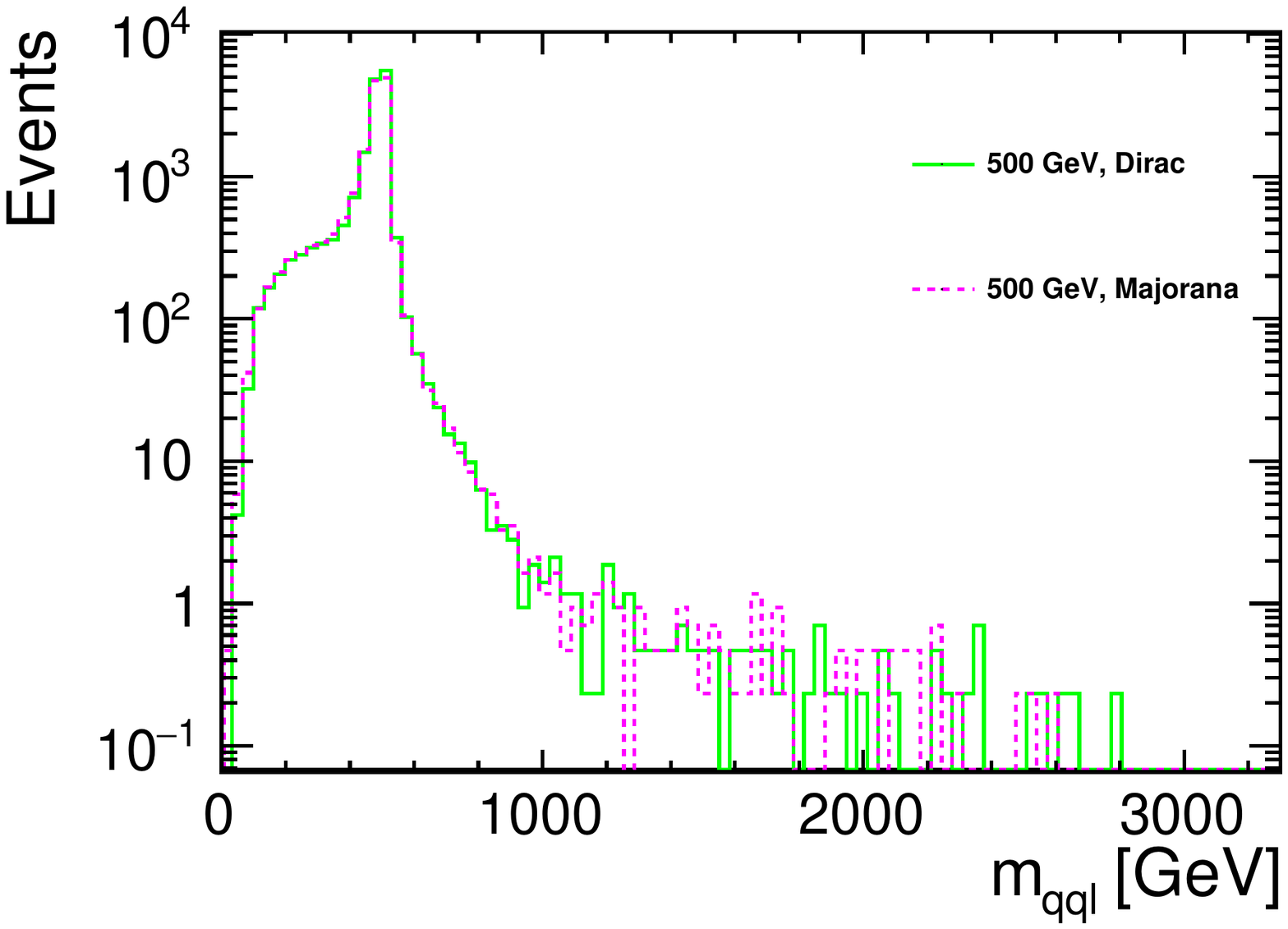}
     \end{subfigure}
    \caption{Distributions of the $E_\ell$ (left) and m$_{qq\ell}$ (right) for the Majorana (pink dashed line) and the Dirac (green solid line) neutrinos with a mass of 500\,GeV at CLIC3000 (detector simulation included).}
    \label{fig:DiMaj_dist}
\end{figure}

Nevertheless, it has to be noted that the kinematic distributions for the Dirac and Majorana cases are not the same. In Figure \ref{fig:DiMaj_angle}, the distribution of the lepton emission angle in the $N$ rest frame at the generator level is shown. The flat distribution for the Majorana neutrino corresponds to the isotropic emission (stemming from an average over the two charge-conjugated decay channels), while for the Dirac case, leptons are emitted mostly in the forward direction. In Figure \ref{fig:DiMaj_dist}, distributions of the lepton energy, $E_\ell$, and reconstructed neutrino mass, m$_{qq\ell}$, at the detector level, are shown. The lepton energy distribution is significantly affected by the different angular distributions, reflecting the neutrino nature. This was first noticed in \cite{Petcov:1984nf}. However, the distributions of the invariant mass of the $qq\ell$ state are almost identical for both cases. As it is the most efficient criterion for the BDT separation, it explains the very similar results for Dirac and Majorana neutrinos.

\begin{figure}
    \centering
    \includegraphics[width=0.8\textwidth]{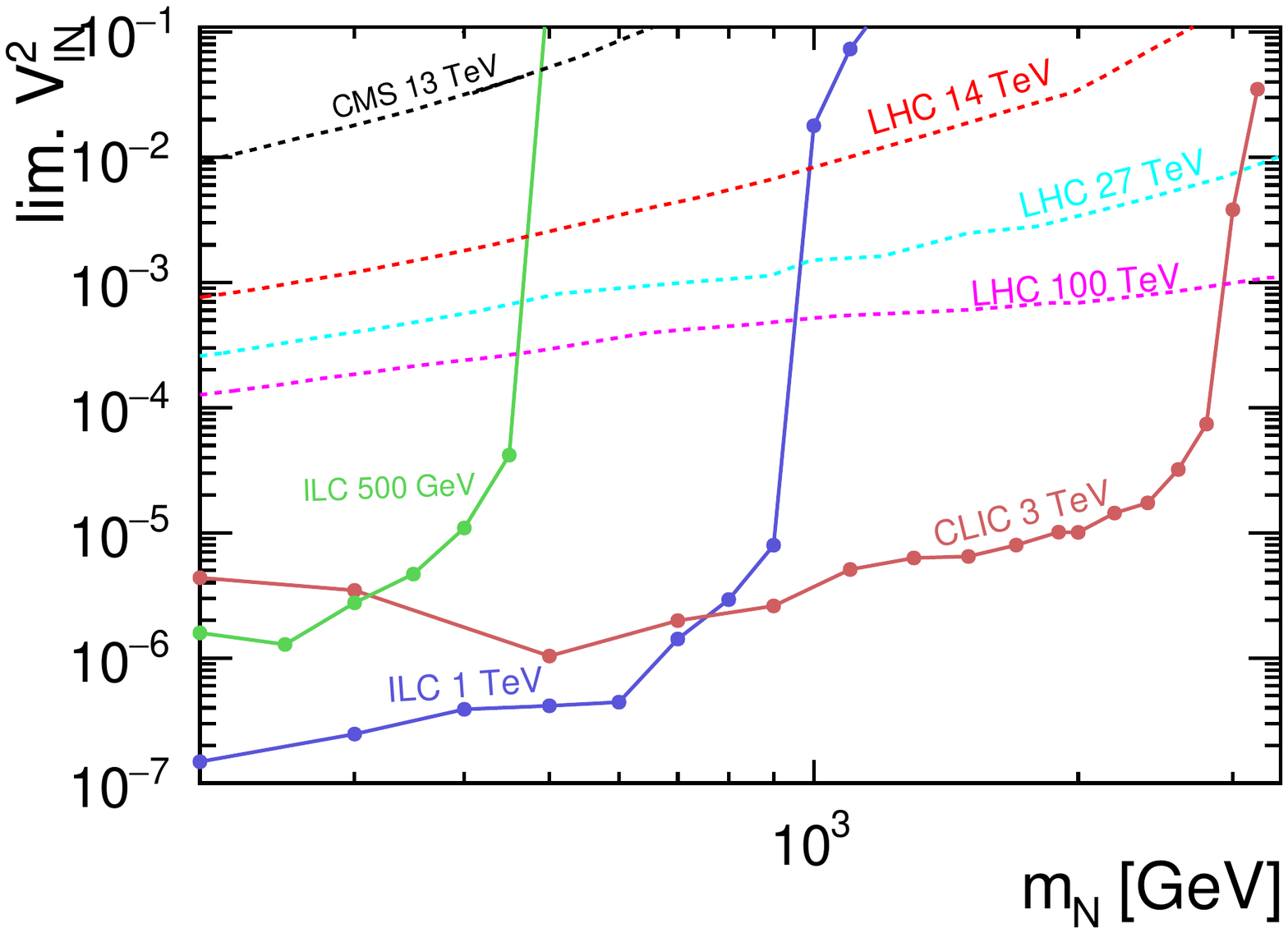}
    \caption{Limits on the coupling $V^2_{\ell N}$ for different collider setups (solid lines: ILC500 -- green, ILC1000 -- violet, CLIC3000 -- dark red). Dashed lines indicate limits from current and future hadron colliders based on \cite{Sirunyan:2018mtv,Pascoli:2018heg}. See the text for details.}
    \label{fig:results}
\end{figure}

The expected limits on the mixing parameter $V_{\ell N}^{2}$ compared to current limits and estimates for future hadron machines are presented in Figure \ref{fig:results}. The limits for the LHC at 13\,TeV come from the CMS Collaboration (Fig. 2 in \cite{Sirunyan:2018mtv}) and were obtained for neutrinos of Majorana nature, while the limits for future high-energy hadron colliders
were taken from the simulation, Fig. 25b, in \cite{Pascoli:2018heg}, where Dirac neutrinos were considered. However, when comparing the results, one should note that in the analyses different assumptions regarding the coupling structure have been made: in hadron collider studies, only two non-zero flavour mixings were taken into account, $V^2_{eN} = V^2_{\mu N} \ne V^2_{\tau N} \equiv 0$, while all the couplings are assumed to have the same non-zero value, $V^2_{eN} = V^2_{\mu N} = V^2_{\tau N}$, in our case. Nevertheless, it was verified that our analysis would give even stronger limits if only two non-zero couplings are considered. It is due to the fact that taus can decay into quarks and then, such events (without electrons or muons in the final state) are excluded from the analysis. On the other hand, as taus can decay leptonically, some of the tau events are included in the analysis, and thus, rerunning of the analysis is needed to compare the results quantitatively with and without employing taus.

\section{Conclusions}
Many theories suggest that, beyond the Standard Model, new particles exist. In some models, these particles are neutral leptons with masses above the electroweak scale which could potentially solve observed cosmological problems, such as the baryon asymmetry or the existence of dark matter. One of the ways to search for such heavy neutrinos could be to use future linear lepton colliders. Nowadays, two concepts of linear lepton colliders are considered: the International Linear Collider (ILC) and the Compact Linear Collider (CLIC). For heavy, weak-scale neutrinos, there are a plethora of different models, depending on whether they address primarily the CP problem of baryogenesis, the dark matter paradigm, or whether they are embedded in theories with extended gauge symmetries like e.g. Grand Unified Theories (GUTs). In this paper, we tried to remain relatively agnostic towards specific models and considered only a single kinematically accessible heavy neutrino species, however, allowing for flavour mixing with all three generations.

Neutrinos of both Dirac and Majorana nature and masses in the range of 200 to 3200\,GeV were considered. We included all relevant $e^+e^- \to X$ SM background processes, as well as those induced by collinear photon splitting from EPA and induced by beamstrahlung photons.
Detector effects were taken into account with the \delphes fast simulation framework.
Making use of multivariate analysis with a BDT classification and the CL$_s$ procedure, we set the exclusion reach of ILC and CLIC for the neutrino mixing parameter $V^2_{\ell N}$, which acts as an effective weak coupling for these heavy neutrinos.
The extracted limits extend down to the coupling values of $V_{\ell N}^2 \lesssim 10^{-7} - 10^{-6}$.
For the heavy neutrino scenarios considered in the presented study, the expected limits are much stricter than the LHC results \cite{Sirunyan:2018mtv} and estimates for the proposed higher-energy hadron machines published so far \cite{Das:2016hof,Pascoli:2018heg}. 
The sensitivity of future $e^+ e^-$ colliders to the heavy-light neutrino mixing is almost insensitive to the neutrino mass up to the production threshold.
Furthermore, for on-shell production of heavy neutrinos, almost the same expected coupling limits are obtained for Dirac and Majorana particles. The variables used in the current study are not yet optimised to distinguish between the Dirac and Majorana hypotheses. Finding an analysis procedure optimised for model discrimination is deferred for future studies.

We note that it might be very interesting to perform a similar study at high-energy muon colliders, which due to their higher anticipated energy of 10 TeV or even beyond could reach much higher neutrino masses. Very mild beamstrahlung will improve the signal-to-background ratio. Also, due to the muon flavour, different flavour mixing structures will be probed. The details are, however, beyond the scope of this paper.

\subsection*{Acknowledgements}

The authors thank Simon Bra{\ss} for technical support with {\sc Whizard}, the UFO model files and the generator-level simulation aspects of this project.
The work was partially supported by the National Science Centre
(Poland) under OPUS research projects no. 2017/25/B/ST2/00496
(2018-2021) and by the Deutsche Forschungsgemeinschaft (DFG, German Research Association) under Germany’s Excellence Strategy-EXC
2121 “Quantum Universe”-39083330.

\bibliographystyle{JHEP}

\bibliography{tex/bibliography}

\end{document}